# The Celestial Sign in the Anglo-Saxon Chronicle in the 770s: Insights on Contemporary Solar Activity

Hisashi Hayakawa (1-2)*, F. Richard Stephenson (3), Yuta Uchikawa (4-5), Yusuke Ebihara (6-7), Christopher J. Scott (8), Matthew N. Wild (1), Julia Wilkinson (9), David M. Willis (1, 10).

(1) Science and Technology Facilities Council, Rutherford Appleton Laboratory, Harwell Campus, Didcot, OX11 0QX, UK
(2) Graduate School of Letters, Osaka University, Toyonaka, Osaka, 5600043, Japan
(3) Department of Physics, University of Durham, Durham, DH1 3LE, UK
(4) University of Tokyo, 7-3-1 Hongo, Bunkyō, Tokyo 1138654, Japan
(5) Fitzwilliam Museum, University of Cambridge, Trumpington St, Cambridge CB2 1RB, UK
(6) Research Institute for Sustainable Humanosphere, Kyoto University, Gokasho, Uji, 6110011, Japan
(7) Unit of Synergetic Studies for Space, Kyoto University, Kitashirakawa-oiwake-cho, Sakyo-ku, Kyoto, 6068306, Japan
(8) Department of Meteorology, University of Reading, Reading, RG6 6BB, UK
(9) Zooniverse, c/o Astrophysics Department, University of Oxford, Oxford, OX1 3RH, UK
(10) Centre for Fusion, Space and Astrophysics, Department of Physics, University of Warwick, Coventry, CV4 7AL, UK

* hisashi.hayakawa@stfc.ac.uk/hayakawa@kwasan.kyoto-u.ac.jp

The Celestial Sign in the Anglo-Saxon Chronicle in the 770s

**Abstract**

The anomalous concentration of radiocarbon in 774/775 attracted intense discussion on its origin, including the possible extreme solar event(s) exceeding any events in observational history. Anticipating such extreme solar events, auroral records were also surveyed in historical documents and those including the red celestial sign after sunset in the Anglo-Saxon Chronicle (ASC) were subjected to consideration. Usoskin *et al.* (2013, *Astron. Astrophys.* **55**, L3: U13) interpreted this record as an aurora and suggested enhanced solar activity around 774/775. Conversely, Neuhäuser and Neuhäuser (2015a, 2015b, *Astron. Nachr.* **336**, 225; **336**, 913: N15a and N15b) interpreted "after sunset" as during sunset or twilight; they considered this sign as a halo display and suggested a solar minimum around 774. However, so far these records have not been discussed in comparison with eyewitness auroral records during the known extreme space-weather events, although they were discussed in relationship with potential extreme events in 774/775. Therefore, we reconstruct the observational details based on the original records in the ASC and philological references, compare them with eyewitness auroral observations during known extreme space-weather events, and consider contemporary solar activity. We clarify the observation was indeed "after sunset", reject the solar halo hypothesis, define the observational time span between 25 March 775 and 25 December 777, and note the parallel halo drawing in 806 in the ASC shown in N15b was not based on the original observation in England. We show examples of eyewitness auroral observations during twilight in known space-weather events, and this celestial sign does not contradict the observational evidence. Accordingly, we consider this event happened after the onset of the event in 774/775, but shows relatively enhanced solar activity, with other historical auroral records in the mid-770s, as also confirmed by the Be data from ice cores.

Hayakawa et al., 2019, *Solar Physics*, preprint.
The Celestial Sign in the Anglo-Saxon Chronicle in the 770s

**1. Introduction**

The anomalous concentration of $^{14}$C in 774/775 in tree-rings (Miyake *et al*., 2012) called for intensive discussions on its origin. Further surveys were conducted on the concentration of $^{14}$C in tree rings around the world and $^{10}$Be in ice cores from the polar regions, to confirm the spike in 774/775 and also another one in 993/994 (Miyake *et al*., 2013, 2015, 2017; Usoskin and Kovaltsov, 2012; Usoskin *et al*., 2013; Jull *et al*., 2014; Mekhaldi *et al*., 2015; Sigl *et al*., 2015; Fogtmann-Schultz *et al*., 2017; Uusitalo *et al*., 2018; Büngten *et al*., 2018). The origin of these spikes attracted various hypotheses, including a nearby supernova (Allen, 2012), a gamma ray burst (Pavlov *et al*., 2013; Hambaryan and Neuhäuser, 2013), a cometary impact with the Earth (Liu *et al*., 2014), and an extreme solar proton event (Thomas *et al*., 2013; Usoskin *et al*., 2013; Miyake *et al*., 2013; Cliver *et al*., 2014).

In the surveys of the tree rings and ice cores, the spikes in 774/775 and 993/994 have been confirmed around the world and in both hemispheres, making a cometary impact and a gamma ray burst unlikely (Usoskin *et al*., 2013; Jull *et al*., 2014; Miyake *et al*., 2015). Mekhaldi *et al*. (2015) surveyed the data on $^{10}$Be and $^{36}$Cl in both Arctic and Antarctic ice cores and detected spikes not only in $^{36}$Cl but also in $^{10}$Be, and concluded that these spikes were caused not by gamma ray bursts but by solar proton events with a hard spectrum, as also suggested by Usoskin *et al*. (2013) and Thomas *et al*. (2013). For the supernova hypothesis, Dee *et al*. (2017) examined the $^{14}$C data around the times of the known seven historical supernovae (e.g., Stephenson and Green, 2002) and showed that none of these supernovae have left distinct signals in carbon-14 data.

Had it been caused by an extreme solar proton event (SPE), it would have yielded much more extreme intensity than any other SPE ever observed to generate these anomalies in cosmogenic nuclides (Smart, Shea, and McCracken, 2006; Cliver and Dietrich, 2013; Cliver *et al*., 2014). This raised a discussion as to whether a superflare or a series of extreme solar events occurred around 774/775, as in other solar-type stars (Maehara *et al*., 2012, 2015, 2017; Shibayama *et al*., 2014; Shibata *et al*., 2013; Notsu *et al*., 2015a, 2015b; Karoff *et al*., 2016; Namekata *et al.*, 2017).

However, this discussion was subject to difficulty caused by the lack of contemporary instrumental observations, as even the telescopic sunspot observations – one of the longest ongoing experiments – had lasted only roughly 400 years (Vaquero, 2007; Vaquero and Vazquez, 2009; Owens, 2013; Clette *et al*., 2014; Vaquero *et al*., 2016).

Therefore, historical records were also surveyed to obtain relevant astronomical evidence, as the historical records have much longer but less complete and homogeneous coverage of datable observations in terms of supernovae (Clark and Stephenson, 1977; Stephenson and Green, 2002;

Enoto, Kisaka, and Shibata, 2018), comets (Kronk, 1999), sunspots (Willis, Easterbrook, and Stephenson, 1980; Vaquero and Vazquez, 2009; Willis and Stephenson, 2001), and aurorae (Silverman, 1998, 2006; Stephenson, Willis, and Hallinan, 2004; Hayakawa *et al.*, 2016b, 2017b).

It is the record of the Anglo-Saxon Chronicle (hereafter, ASC) that provided one of the first insights for this discussion. Allen (2012) suggested the celestial sign of a "red crucifix" in the ASC would have been a supernova in 774. Gibbons and Werner (2013) extended the survey of historical documents to add another record from Germany and redate the celestial sign in the ASC to 776. However, the supernova hypothesis seems unlikely due to the lack of supernova remnants and contemporary observations (Stephenson and Green, 2002; Usoskin *et al.*, 2013).

Partially based on the existing auroral catalogues (Link, 1962; Yau, Stephenson, and Willis, 1995), Usoskin *et al.* (2013, hereafter U13) surveyed further auroral records around 775 and suggested enhanced solar activity around 775, based on the concentration of auroral records around 775, including the celestial sign in 773/774, mentioning its redating to 776 by Swanton (2000). Stephenson (2015, hereafter S15) revisited the critical editions of the original historical sources and clarified that their reliability should be carefully considered. S15 concluded that there is one "definite" auroral record in China (see, Stephenson et al., 2019, for further details), dated on 12 January 776, and one plausible record in the ASC, for which the observational year is in doubt.

Immediately after S15, Neuhäuser and Neuhäuser (2015a, hereafter N15a) made another critical survey based on historical documents, just as done in S15. N15a proposed five criteria for the likeliness of "strong aurora" and suggested that the celestial sign of a "red cross" or "red crucifix" in the ASC was not an auroral display but a halo effect and should be dated in 776 based on its later critical edition (Garmonsway, 1953). Their follow-up paper (Neuhäuser and Neuhäuser, 2015b, hereafter N15b) revisited the records of the ASC based on its manuscripts and critical editions and carried out further discussions on this topic.

Thus, the physical nature of the celestial sign described in the ASC around 774/775 has been the topic of discussions with different interpretations: supernova (Allen, 2012), aurora (U13 and S15), and halo (N15a and N15b). Such discussions are important to understand the background solar activity around the cosmic ray event in 775. U13 identified enhanced solar activity around 774/775, based on the auroral candidates including this celestial sign. S15 and Cliver *et al.* (2014) considered this celestial sign and another Chinese record as aurorae but suggested caution in judging the level of associated solar activity. Independently, Sukhodolov *et al.* (2017) examined the Be data in ice cores and estimated the maximum of the solar cycle as around 776 -- 777 as shown in their Figure 1. On the other hand, N15a and N15b remarked on the absence of auroral reports from 774 to 785 and placed the minimum of the sunspot cycle around 774±1 (N15a, pp. 244-245).

Therefore, it is important to reconsider the physical nature of this celestial sign to understand the background solar activity around 774/775. The initial difficulty is partially caused by the rather fragmentary nature and limited amount of information on historical records in this period (Silverman, 1992, 1998; Barnard *et al.*, 2017). In this sense, it is always important to analyze the given records in comparison with philological results based on the original records, and parallel cases of visual observations in early-modern times, to derive reasonable conclusions, as is also done for the reconstruction of the amplitude of the Maunder Minimum (Zolotova and Ponyavin, 2015; Usoskin *et al.*, 2015; Vaquero *et al.*, 2015), the reliability of sunspot drawings (Zolotova and Ponyavin, 2015; Usoskin *et al.*, 2015; Carrasco, Álvarez, and Vaquero, 2015; Carrasco, Vaquero, and Gallego, 2018), the reliability of auroral records at that time (Zolotova and Ponyavin, 2016; Usoskin *et al.*, 2017), and intensity of magnetic storms in the 19 -- 20 centuries (Silverman and Cliver, 2001; Cliver and Dietrich, 2013; Silverman, 2008; Hayakawa *et al.*, 2018f).

The article aims at revisiting the variants of this celestial sign and analyzes them philologically, based on contemporary records and modern philological results, and compares them with early-modern observations and modern knowledge of auroral displays and halo displays. The aim is to provide some insights into the amplitude of solar activity around the cosmic-ray event in 775.

## 2. Method

In order to understand the physical nature and the timing of the phenomenon under investigation as far as possible, we first clarify the source documents and original records, based on the latest critical editions of each manuscript. We consider the background to these manuscripts and discuss a possible range of observational sites.

Then, we consider the available records and descriptions to clarify the shape and the observational timing of this celestial sign, as they are important clues to distinguish between the auroral hypothesis and the halo hypothesis. The observational timing is described as "after sunset" and this wording was literally adopted by U13 and S15 with an auroral hypothesis. On the other hand, this was interpreted as "during sunset" or "during twilight" by N15a and N15b, leading to the halo hypothesis. Here, we consult the dictionaries of Old English and Medieval Latin and compare their contents with the parallel contemporary records. We then reconsider the observational year for which U13 showed 773/774 from an existing catalogue and 776 from the redating of a critical edition, S15 set a caveat on its precise year of occurrence, and N15a and N15b suggested to date it in 776 based on a "two-year shift" in a critical editing (Garmonsway, 1953).

Based on these philological analyses, we compare this celestial record with modern knowledge and early-modern observations of halo displays and auroral displays. In this context, we reconsider the relevant "likeliness criteria" for auroral candidates recently suggested by N15a, with reference to future scientific discussions upon the reliability of historical auroral candidates.

**3. Understanding Source Documents**

**3.1. Manuscripts and Records**

As explained in Swanton (2000, pp. xxi-xxviii), ASC is a term employed by modern scholars to a composite set of Old English annals recording the events that form the basis of our understanding of the Anglo-Saxon history. The original compilation known as the "common stock" was probably written in the court of King Alfred of the West-Saxons in the early 890s, while it is considered lost. Copies of this "common stock" were sent to religious houses and later events were added to these copies by various hands at different times with their own interests. So far, seven manuscripts and one fragment are currently known (see Table 1). The ASC was partially translated and incorporated to the Latin chronicles such as those by Asser, Æthelweard, Henry of Huntingdon, and John of Worcester.

Table 1: Manuscripts of the ASC with their date, archive, and language(s). OE and ML mean Old English and Medieval Latin respectively. See Swanton (2000, pp. xxix) for the tree for the relevant manuscripts and associated chronicles. Here, "*c.*" denoted "*circa*" in terms of chronology and "*f.*" denotes "*folio*" in terms of its foliation.

|   | Manuscript | Date | Archive | Language |
|---|---|---|---|---|
| A | Cambridge, Corpus Christi College MS 173, *ff*. 1v-32r | *c*. 891 | Winchester | OE |
| B | British Library MSS Cotton Tiberius A iii, f. 178; A vi, *ff*. 1-34 | 977-979 | Abingdon | OE |
| C | British Library MS Cotton Tiberius B i, *ff*. 115v-64 | mid-11th cen. | Abingdon | OE |
| D | British Library MS Cotton Tiberius B iv, *ff*. 3-86 | mid-11th cen. | Worcester | OE |
| E | Oxford, Bodleian Library MS Laud 636 | 1116-1121 | Peterborough | OE |

| F | British Library MS Cotton Domitian A viii, *ff.* 30-70 | *c.* 1100 | Canterbury | OE/ML |
| A[2]/G/W | British Library MS Cotton Otho Bxi, 2 | 1001-1012/13 | Winchester | OE |
| H | British Library MS Cotton Domitian A ix, *f.* 9 | *c.* 1114 | Winchester | OE |

The record for the celestial sign in the 770s (*e.g.* Figure 1) is included in seven of these manuscripts, while MS A²/G/W is a copy of MS A and will not be considered here (Swanton, 2000; N15b). Transcriptions and translations of the relevant texts are shown in Appendix 1, based on the Collaborative Editions (Bately, 1986; Taylor, 1983; O'Brien O'Keeffe, 2001; Cubbin, 1996; Irvine, 2004; Baker, 2000) and the translation by Swanton (2000).

Figure 1: The record of the celestial sign in ASC MS A (*f.* 10v), the earliest surviving manuscript of the ASC (c) Corpus Christi College, Cambridge (MS 173, *f.* 10v).

As shown in Table 1, five manuscripts (MS A-E) are in Old English and one manuscript (MS F) is bilingual in Old English and Latin. While N15b also showed their transcriptions and translations, it should be noted that their transcriptions are not consistent with the known letter system of Old English (Marsden, 2004, pp. xxix-xxxiv). N15b systematically rendered "þ" as "b" and "ð" as "o" from the known transcriptions (Bately, 1986; Taylor, 1983; O'Brien O'Keeffe, 2001; Cubbin, 1996; Irvine, 2004; Baker, 2000; Appendix 1). However, the letters "þ (thorn)" and "ð (eth)" are not variants of "b" and "o" but representatives for dental fricative and derived from the runic letter and modified roman letter (*e.g.* Freeborn, 1998, p.24; Marsden, 2004, p. xxix; Shaw, 2013).

Note that we use the term "Old English" for the language in Anglo-Saxon England in this paper, while N15a and N15b use the term "Medieval Anglo-Saxon" for this language. The *Willey Blackwell Encyclopedia of Anglo-Saxon England* (Lapidge *et al.*, 2014, pp. 350-352) explains that Old English, the language of the Anglo-Saxons, is "sometimes called Anglo-Saxon, though that term is now outmoded".

### 3.2. Observational Area

The observational site of this celestial sign is not described clearly in the ASC. However,

considering that the celestial phenomenon was followed by the description of the battle of Otford fought between the two Anglo-Saxon kingdoms, *i.e.* between the Mercians and the Kentish, as if the two events were related, it would be conservative to consider that the sign was observed in England (see Section 5.1 for the case of another celestial sign in 806). Whitelock (1979, p. 126) considers the records between 755 (*recte* 757) and 828 (*recte* 830) in the ASC are based on reports in Mercia, while Swanton (2000, p. xviii) considers the original "common stock" of the ASC was compiled in the western part of Wessex such as Dorset or Somerset.

Accordingly, in order to narrow down the possible observational area for the sake of convenience, we set the political and religious centres of major kingdoms -- York (Northumbria), Winchester (Wessex), Canterbury (Kent), and Worcester (Mercia) -- as the references for the possible area of observation (see Table 2). Note that the contemporary courts were moving with the itinerant kings in medieval Western Europe and their locations were not necessarily the same as these centres (Peyer, 1964; Bernhardt, 1993; Roach, 2013).

Assuming the dipole model of CALS3K.4b (Korte and Constable, 2011), we compute the location of the magnetic north pole as N86.7° to N86.8° in latitude and E78.9° to E79.2° in longitude from 773 to 777. Calculating the magnetic latitude (MLAT) based on the angular distance of given locations and the magnetic north pole, we compute the MLAT of these sites to be in the range 54° to 52° in MLAT, as shown in Table 2.

Table 2: Location of centres in Northumbria, Wessex, Kent, and Mercia with their geographical coordinates and approximate magnetic latitudes (MLAT) from 773 to 777, according to CALS3K.4b.

| Kingdom | Site | Latitude | Longitude | MLAT |
|---|---|---|---|---|
| Northumbria | York | N53°58′ | W001°05′ | ≈ 54.3 -- 54.5 |
| Wessex | Winchester | N51°04′ | W001°19′ | ≈ 51.4 -- 51.6 |
| Kent | Canterbury | N51°17′ | E001°05′ | ≈ 51.8 -- 51.9 |
| Mercia | Worcester | N52°12′ | W002°13′ | ≈ 52.5 -- 52.6 |

### 4. Understanding the Celestial Sign

#### 4.1. Shape

The shape of the celestial sign is described as a "red crucifix" (Allen, 2012; Gibbons and Werner, 2012; N15a; N15b) or a "red cross" (U13; S15; N15a; N15b), while U13 notes another translation as "red sign of Christ" citing Swanton (2000).

These differences are caused by the variation in the Old English manuscripts and the Medieval Latin translations of the ASC (see Table 1). As shown in Appendices 1 and 2, this celestial sign is described as "*read Cristes mæl/mel* (red sign of Christ)" in the Old English variants (MS A – MS F) and translated as "*signum dominicæ crucis* (sign of the Lord's cross)" or "*crucis signum* (sign of the cross)" in the Medieval Latin variants after the late 10th century (MS F and Chronicle of Æthelweard). In the process of translation, mention of the colour had disappeared and the Lord's sign had been specified as a cross. Twelfth-century sources rendered it differently; John of Worcester (Darlington, McGurk, and Bray, 1995, pp. 210-211) described it in his chronicle as "*rubicundi coloris signum in crucis modum* (a red sign after the fashion of a cross)", combining the Old English and previous Latin versions, while in the History of the English (*Historia Anglorum*; Greenaway 1996, pp.250-251) by Henry of Huntingdon, it was merely referred to as "*rubea signa* (red signs)" with the change of number from singular to plural.

From these records arise several possibilities concerning the shape of this celestial sign. Firstly, the term "sign of Christ" does not necessarily denote the "sign of the cross" but also baptism and the stigmata (the wounds Christ suffered on the cross) (Lenker, 2010, p. 263; Bremmer, 2010, pp. 205-214). Moreover, evidence in philology and archaeology shows that Anglo-Saxons recognized crosses with various shapes as well. The contemporary coins from Anglo-Saxon England show not only a vertical cross but also XP (chi-rho, ☧), TP (tau-rho, staurogram), IX (✶) (Bremmer 2010, p. 205), a diagonal cross, and a cross with wedges in angles (see, Figure 2 and Appendix 3; see also, Naismith 2017, Plates 33-38).

Therefore, we respect the original forms of the Old English and Latin terms and translate them literally. Also it is more conservative not to determine the shape of this celestial sign as a vertical cross as done in previous studies but to consider this celestial sign from wider possibilities such as diagonal crosses and Chi-Rho (☧), as in Figure 2.

Hayakawa et al., 2019, *Solar Physics*, preprint.
The Celestial Sign in the Anglo-Saxon Chronicle in the 770s

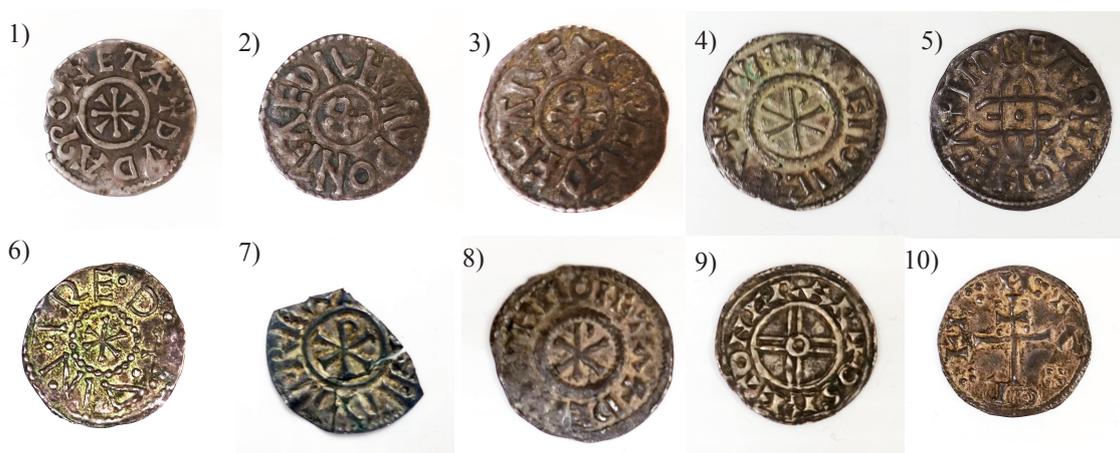

Figure 2: Signs of Christ in Anglo-Saxon coins with a vertical cross, a cross with wedges in angles, chi-rho, tau-rho, etc. Their classification by Naismith (2017) is shown at Table 3 (by courtesy of Fitzwilliam Museum, University of Cambridge; see also Naismith, 2017). The identifications (IDs) in Table 3 are given from left to right in the first row (1-5) and the second row (6-10) in this figure.

Table 3: The detail of coins in Figure 2 provided by Naismith (2017; N17). The period shows when the very coin was probably produced. In this table, the description "*c*. 800/10 – 21" indicates its chronological range starts approximately in 800 or 810 and ends in 821.

| ID | Issuer | Mint | Period | Description of Sings of Cross/Christ | No. in N17 |
|---|---|---|---|---|---|
| 1 | Cuthred, king of Kent (*r*. 798-807) | Canterbury | *c*. 805 -- 7 | cross pattée with wedges in angles | 960r |
| 2 | Æthelheard, archbishop of Canterbury (*r*.792-805) | Canterbury | 792/3 -- 6 | staurogram (tau-rho) | 970o |
| 3 | Æthelheard, archbishop of Canterbury (*r*.792-805) | Canterbury | 792/3 -- 6 | staurogram (tau-rho) | 970r |
| 4 | Ceolnoth, archbishop of Canterbury (*r*. 833-70) | Canterbury? | *c*. 844 -- 9 | chi-rho | 984r |
| 5 | Coenwulf, king of the Mercians (*r*. 796-821) | Canterbury | *c*. 810 -- 21 | pincer cross with wedges in angles | 1060r |
| 6 | Coenwulf, king of the | East Anglia | *c*. 800/10 -- | hexagram (IX monogram?) | 1066r |
|    | Mercians (*r*. 796-821)                      | (Ipswich?)  | 21                 |                                              |       |
| 7  | Beorhtwulf, king of the Mercians (*r*. 840-52) | London?    | 840 -- 52          | chi-rho                                      | 1094o |
| 8  | Æthelwulf, king of the West Saxons (*r*. 839-58) | Canterbury? | *c*. 844 -- 9    | chi-rho                                      | 1099r |
| 9  | Cnut, king of the English (*r*. 1016-35)     | Cambridge   | *c*. 1029 -- 1035/6 | voided short cross with annulet at centre   | 2039r |
| 10 | Cnut, Viking king of Northumbria             | York        | *c*. 895 -- 905    | patriarchal cross                            | 2531r |

**4.2. After Sunset or During Sunset**

As reviewed previously, the meaning of "after sunset" was one of the main foci of discussions on the physical nature of this celestial sign. On the one hand, U13 and S15 considered "after sunset" literally as after sunset, while S15 noted "it was not stated that at the time the sky was dark" and suggested the possibility of appearance in the twilight sky.

On the other hand, N15 considered this celestial sign in the ASC as a halo display, because: (1) the timing after sunset usually means twilight, (2) in Old English (and Latin) the word after or *æfter* (in Latin *post*) can also mean during (sunset). Accordingly, N15a (p.239) interpreted this event as "a solar halo display at least with horizontal arc and vertical pillar, together apparently looking like a Red Cross during sunset" if it is "during" sunset or "lunar halo display" if it is "after" sunset.

The discussion by N15b seems partially based on their criterion "(iv) Night-time observation (darkness)" in N15a. Here, N15a considered "wordings like *after sunset* or *before sunrise* would not necessarily indicate aurorae, but also do not disprove the possibility of an aurora, so that they are considered "neutral" while "If the event clearly happened during civil or nautical twilight, it cannot be concluded that it was an aurora".

It is important to understand precisely the original manuscript, as N15b suggested. However, their interpretation of the preposition "*æfter*" in Old English and "*post*" in Latin is slightly puzzling. We therefore start our analysis with the Latin variants of the Anglo Saxon Chronicle.

While all manuscripts provide Old English text, the bilingual MS F and the Chronicle of Æthelweard[1] provide Latin text (see Appendices 1 -- 2). As shown in Appendix 1, the MS F (*f*.

---

[1] A late 10th century Latin chronicle which made use of an early version of the ASC (see Figure 1).

48v.), the Chronicle of Æthelweard (Campbell, 1962, pp. 25-26; see Appendix 2), and later Latin works describe the timing of the "red cross" either *post solis occubitum*, *post solis occasum*, or *post occasum solis*. The words *occubitum* and *occasum* are singular masculine accusative of the nouns *occubitus* and *occasus* with the meaning of a going down, setting (eccl. Lat.) and a falling, going down, setting (of the heavenly bodies), both of which are used with *solis* to mean sunset (Lewis and Short, 1879, pp. 1250--1251).

It is therefore the meaning of *post* which determines the timing of this event in relation to sunset. As we have already mentioned, the Latin word *post* usually means not "during/around" but "*Of time*, after, since" or "at a time subsequent to, after; for or in the period subsequent to, since" as given in dictionaries for classic Latin such as Lewis and Short (1879, p.1404) and Glare (1982, p.1412). What we need to consider is the use of *post* in Medieval Latin, especially in England.

While N15a and N15b state "in Old English (and Latin) the word after or *æfter* (in Latin *post*) can also mean during (sunset)" (N15a, p. 239) and "Both the English *after/æfter* (Clark Hall 1960) and the Latin post (Niermeyer 1976) had the meaning of both our todays after as well as during/around [*sic*]" (N15b, p.918) respectively, their interpretation of Latin *post* is not attested by the dictionary they cite (Niermeyer, 1976).

Moreover, while N15a and N15b interpret *æfter* in Old English as "during" based on Clark Hall (1960), this interpretation is not taking the Latin translation into account. As reviewed by Fulk (2009) and Rodríguez (2018), Clark Hall's dictionary is designed for learners, contrary to the scholarly dictionaries such as Bosworth-Toller and the Dictionary of Old English (DOE). Bosworth and Toller (1921, p. 10) assigned the meaning of *during* only in a pair of *per* in Latin, not that of *post* in Latin in the *local and temporal* context. The interpretation of during is not registered in the DOE, the latest and most comprehensive dictionary for Old English[2]. Therefore, the latest philological results based on Old English do not favour the interpretation "during sunset" but "after sunset". In short, the philological results tell us to interpret *æfter sunnan setlgonge/setlgange* in Old English or *post solis occubitum/occasum* in Medieval Latin as "after sunset" literally and not to manipulate them to mean "during sunset". These details are shown in the Appendix 4.1.

**4.3. During Twilight or Dark Night**

While we can define the observational time of this celestial sign as indeed "after" sunset, the variants of the ASC do not tell us much about its further timing, as S15 suggested. What are defined in the dictionaries are the interpretations of both *æfter* in Old English and post in Latin as after or subsequent to some event or timing. We need to consider if the term after sunset restricts timing to during twilight as speculated in N15a (p. 239) and hence does not satisfy one of their criteria: "night-time observation (darkness)" (N15a, p. 230).

---

[2] *The Dictionary of Old English: A to I Online*, s.v. *æfter* (https://tapor.library.utoronto.ca/doe/; accessed: 17 November 2018)

However, first of all, auroras are reported even "after sunset" when they are bright enough. For example, "Indications of the aurora were noticed here soon after sunset" at Glasgow (N54°41′, W001°13′; 57.8° MLAT) on 4 February 1872 (Bottomley, 1872, p. 326), when auroras were seen globally (Silverman, 2008; Hayakawa *et al*., 2018d).

Moreover, the parallel records in contemporary time may provide some insights on this topic. The Alfred's Law Code is one of the most extensive legitimate enactments surviving from Anglo-Saxon England, compiled during the reign of King Alfred (871 -- 899) (Dammery, 1991). In this code, we find a case where "after sunrise" is used as an opposite to by night, to indicate that these terms mean daytime and night-time, respectively (see Appendix 4.2).

These examples provide counter-examples against the interpretation by N15a (p.239) who restrict the timing of "after sunset" as during twilight and rejected the possibility of an auroral observation. While we do not exclude the possibility that this celestial sign was indeed observed during twilight after sunset, we should not exclude the possibility that it was observed in the dark night after twilight, either. Moreover, "twilight" is generally described as *deorcung*, *glóm*, or *glómung* in the Old English texts (Bosworth and Toller, 1921, p. 202 and p. 481).

### 4.4. Year of Observation

Defining the year of observation is subject to difficulty because of the variety of years in the variants of the ASC. U13 showed its year as 773/774, citing Link (1962), but showed the redating to 776, according to the latest critical edition (Swanton, 2000). S15 consulted a classic critical edition (Longmans and Roberts, 1858) casting a caveat on its dating with variants of 773, 774, and 776 in its versions: "There is clearly some doubt about the precise year of occurrence" (S15, p. 1542). N15a and N15b consulted a later critical edition (Garmonsway, 1953) and suggested a date in 776, based on a two-year shift in the critical editing. The timing was discussed intensively, partially to determine if this event occurred before or after the anomalous concentration of $^{14}$C in 775.

The reason for the variety in the observational year has been caused by the variants as discussed in S15 and N15b. As shown in Appendices 1 and 2, MS A dated this event as 773 and MSS B to F dated it as 774, whereas the Chronicle of Æthelweard dated it as 772. As suggested in U13 and N15b, the critical editions generally redate this event as 776 as in Garmonsway (1953) cited in N15b, or in Swanton (2000) cited in U13. The critical editions (Swanton, 2000, p. 46; Garmonsway, 1953; Whitelock, 1961, p.30) have noted the chronological dislocation during the interval 754 to 845 and have considered that the records are dated two to three years earlier than in reality.

This is confirmed by the chronology of common events in independent contemporary chronicles such as Bede's *Ecclesiastical History of the English People*, *Gesta Veterum Northanhymbrorum* ("Northern Annals") in *Historia Regum*, or other continental chronicles (Stubbs, 1868, pp. xc-xcv). The multiple examples shown in Stubbs (1868, pp. xci-c) would be sufficient to confirm the two-year backward shift in the variants of the ASC such as the deaths of Pope Adrian (794 in ASC *vs*. 25 December 795), of Charles the Great (812 in ASC *vs*. 28 January 814), and of Leo III (814 in ASC *vs*. 12 June 816). The Annals of St. Neots (compiled in the second quarter of the 12th century), a Latin chronicle based on the early variants, show their chronology displaced two years after those of other variants.

We have to be aware that the onsets of the year in the ASC vary between modern new year (1 January), Annunciation (25 March), Roman Indiction (1 September), and Christmas (25 December) (Swanton 2000, pp. xv--xvi). Swanton (2000, p. xvi) notes "[c]learly, events ascribed to a year beginning either in September or in March might well be dated a year too early or a year too late by modern reckoning beginning 1 January". This explains the variation of chronology of the celestial sign in the ASC (see also N15b) and gives us up to a one-year uncertainty back and forth for the timing of this celestial sign (25 March 775 – 25 December 777).

## 5. Comparison with Parallel Scientific Observations
### 5.1. Halo Hypothesis after Sunset

As clarified above, this celestial sign is not necessarily defined as a vertical red cross, as considered in previous studies (Section 4.1). Moreover, even if we consider the case of a vertical cross for this event, it seems the halo hypothesis by N15a and N15b is not consistent with the descriptions in the ASC.

First of all, philological results tell us that both *æfter* in Old English and *post* in Latin do not mean during but after and hence the description of after sunset should be understood literally (Section 4.2). This insight is important when considering the halo hypothesis in N15a and N15b, who interpreted this celestial sign as a "red cross", namely a combination of "the horizontal arc and a vertical pillar of light (either with the Sun during sunset or with the moon after sunset)" (N15b, p. 913), or "a solar halo display at least with horizontal arc and vertical pillar, together apparently looking like a red cross during sunset" (N15a, p. 239).

However, after sunset, the Sun is below the horizon and the horizontal arc cannot be seen, as shown in Figure 3. Even if the vertical light pillar were seen above the horizon at this time, it would not form a red cross just by itself (see also Minnaert, 1993, pp. 219--223). Therefore, we reject the solar halo hypothesis by N15a and N15b for this red cross due to the timing of after sunset.

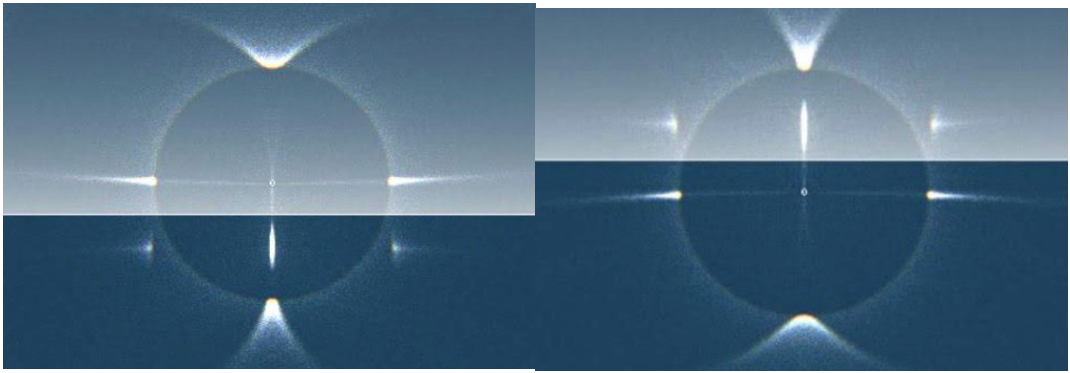

Figure 3: The solar halos before sunset (left: 6° above the horizon) and after sunset (right: 6° below the horizon), generated by HaloSim 3 based on Tape (1994). This software is available at https://www.atoptics.co.uk/halo/halfeat.htm. The position of the solar disk is shown by the small white dot. The pale blue part and the dark blue parts correspond with the sky above the horizon and the part hidden under horizon. As shown here, the solar halo cannot produce a "red cross" as combination of a "horizontal arc and a vertical pillar of light" (suggested by N15a and N15b) after sunset, as the horizontal arc is below the horizon.

Another possibility suggested by N15a and N15b is the combination of a "horizontal arc and a vertical pillar of light" in the case of the lunar halo. It should be noted that this red cross was reported specifically with a red colour. Minnaert (1993, p. 209) explains in his textbook "Most of those [N. B. halo display besides the small halos] that have been seen were caused by the sun; those belonging to the moon are much fainter and their colors are practically imperceptible", as only the rod cells are at work under the illumination of moonlight and do not provide impression of colour but that of light (Minnaert, 1993, p. 133). Therefore, it seems difficult to expect the red colour for the lunar halo, according to Minnaert (1993).

N15b (p. 921) provides a drawing of a paraselene on 24 January 1680 [corresponding to 3 February 1680 in the Gregorian calendar] by Gottfried Kirch for their justification of relating a red cross with a lunar halo. However, according to the letter from Gottfried Kirch, it was neither a vertical pillar (h-k in Kirch's annotations) nor horizontal arc (g-c-b-f in Kirch's annotations) but mock-moons (c and b in Kirch's annotations) and the arc near the zenith (n-l-m in Kirch's annotations) that had colours like a rainbow (Paris Observatory MS C 1/16; Herbst, 2006, p. 285). The vertical pillar is situated along the border of *f.* 2299v and *f.* 2300r in the folios (f.) of Kirch's correspondence collection (Paris Observatory MS C 1/16) and partially affected by the apparent colour caused by the border of these folia, as the original manuscript MS C1/16 shows. In this case, even if the mock-moons and arc near the zenith were coloured, they cannot be a "red cross", as neither the horizontal arc nor the vertical pillar are coloured in red.

Figure 4: A parallel record of a cross-sign on 4 June 806, denoting a lunar halo (c) the British Library Board (MS F, *f.* 51r), the same cross-sign in an autograph manuscript of *Annales Laudunenses et S. Vincentii Mettensis breves* (c) the Staatsbibliothek zu Berlin (MS Phillipps 1830, *f.* 7v), and transcriptions of *Annales Sancti Maximi Trevirensis and Annales Laudunenses et S. Vincentii Mettensis breves* in MGH (Pertz, 1841, p. 6; Pertz, 1888, p. 1294).

Interestingly, while N15b showed a drawing of a cross in MS F on 4 June 806 and mentioned another record of a solar halo, these parallel records show that the Moon or the Sun is explicitly mentioned in the case of solar and lunar halos (see Appendix 5.1). This halo record is even described using the different terminology "*rodetacn* (cross-sign)" in the Old English version (MS F, f.51r.) from the other records of a "red cross" in 776. Furthermore, as shown in Figure 4, this halo drawing is not based on the original observation in Anglo-Saxon England but a copy from those in European continental chronicles in mid-9th century (Baker, 2000, pp. xlvi-xlvii, 59). Therefore, it is not likely that the records in 776 and 806 are the same phenomenon, while they contain the same term "sign of the cross (*signum* (*dominicæ*) *crucis* / *crucis signum*)" in the Latin version: the former due to the translation from Old English and the latter the translation to Old English (see Appendix 5.2).

In short, we reject the solar halo hypothesis by N15a and N15b based on the philological interpretation of its observational time as "after sunset", as the horizontal arc would have been below the horizon and would not have been visible to the observer at that time.

**5.2. Auroral Displays like this Celestial Sign**

Now that the solar halo hypothesis by N15a and N15b is rejected and the lunar halo hypothesis by N15a and N15b appears unlikely, the auroral hypothesis (U13 and S15) becomes rather more persuasive again. This is also supported by the fact that "after sunset" is not necessarily confined to the interval just after the sunset, namely during twilight.

Whatever the "red sign of Christ" means, we need to be aware that the Bible was a key reference for natural phenomena observed in Medieval Europe (*e.g.*, Silverman, 1998; U13) and it is possible that the description conformed to religious signs. The form of this celestial sign is even not necessarily confined to be a vertical red cross.

During the early modern extreme space weather event on 1872 February 4 (*e.g.*, Silverman, 2008; Hayakawa *et al*., 2018d), an observer at Bodmin (N50°28′, W004°43′; 54.5° MLAT) compared a "magnificent display of aurora between 6 and 8 p.m. (on the 4th)" with "a beautiful Maltese cross" when he observed a convergence of streamers just below Capella (MMM Editorial, 1872, p. 32) with its altitude calculated 74°21′.

The cross with wedges in angles is reproduced by a corona aurora. Corona auroras are rayed auroras seen around the magnetic zenith (Chamberlain, 1961, pp. 117-118). Figure 5 shows an auroral display observed by Hall (1872) at Barnstaple (N51°05′, W004°04′; 55.0° MLAT) at 18:55 LT on 4 February 1872, when "Outburst of rays in every part of the sky, all diverging from a point 10° E. of the Pleiades, where the ends of the rays are very distinctly seen to interlace each other for a length of about 4°, producing a great intensity of light" (Hall, 1872, p. 2).

This structure resembling a combined-cross can be reconstructed by considering auroral emissions along the dipole magnetic field lines at $L$=2.6~4.6 within the height range 150-800 km. The longitudinal range of the magnetic field lines drawn is arbitrarily chosen. The location of the dipole axis is based on the GUFM1 magnetic-field model (Jackson et al., 2000).

Moreover, even the form of cross, TP (tau-rho), or XP (chi-rho) can be reproduced by a corona aurora, or a combination of a horizontally-extended auroral arc and a vertically-extended ray structure. Lemström (1886, pp.10-11) reported an observation of a corona aurora at Enare (*i.e*. Inari) on 1871 November 16. It was in a "bizarre form". It represented a "folded red veil in a loop" with its convexity at the zenith as shown in Figure 6.

This ribbon-like structure resembling TP or XP can be reconstructed by considering auroral emissions along the dipole magnetic field line at $L \approx 6.4$ for the two tips, and the dipole magnetic field lines at $L \approx 5.6$ -- 6.1 for the loop. The height range between 100 and 150 km (possibly corresponding to the green emission) is drawn in green, and the range between 150 and 400 km (possibly corresponding to the red emission) is drawn in red. The longitudinal range of the magnetic field lines drawn is again arbitrarily chosen. The geomagnetic pole, at which the dipole axis intersects the surface of the Earth, was located at N78.6° and E292.6°, as determined by the GUFM1 model (Jackson *et al*., 2000). The emission around the centre of the auroral corona was probably low due to some reason, which may have shown this auroral display like a loop.

Sometimes, the horizontally-extended auroral arcs and the vertically-extended ray structures can been seen as if they intersect each other. Consequently, it may appear to be a cross-like structure. One possible configuration is that the tall ray structures are located far from an observer and that the auroral arcs are located in the vicinity of the observer. One of the classes of tall ray structures is a sunlit aurora that can result from resonant scattering by $N_2^+$ in sunlit conditions (Størmer, 1955, pp. 91-92; Hunten, 2003). The tall ray structures can be seen from the ground in darkness. Another class of the tall ray structures is associated with ohmic heating in field-aligned filaments (Otto *et al.*, 2003).

Figure 7 shows an auroral drawing with a few cross-like structures by Capron (1879, p. 22) dated on 4 February 1874. These crosses were described in the following terms: "fleecy horizontal clouds of misty light floated in the north above the bow across the streamers". Capron made spectroscopic observations of this display and found auroral spectra in it. He described it as follows: "I found Ångstrom's line quite bright, and by the side of it three faint and misty bands towards the blue end of the spectrum upon a faintly illuminated ground. I could also see at times a bright line beyond the bands towards the violet".

This cross-like structure is reconstructed by considering auroral emissions along the dipole magnetic field lines at $L$=4.0 within the height range 100 -- 200 km and at $L$ = 5.0 within the height range 150 -- 600 km. The geomagnetic pole, at which the dipole axis intersects the surface of the Earth, was again determined by the GUFM1 model (Jackson *et al.*, 2000).

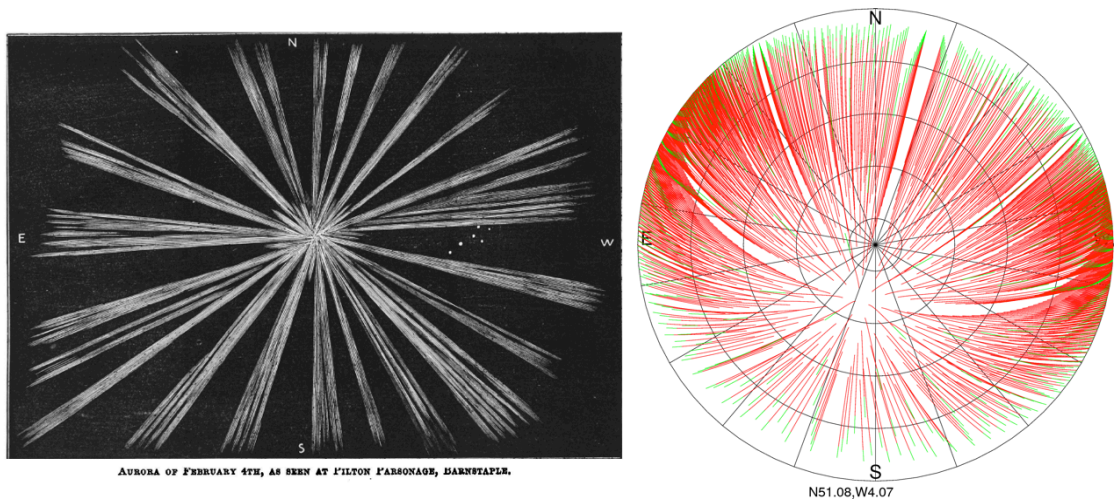

Figure 5: (left) The auroral drawing on 4 February 1872 by Hall (1872); (right) our reconstruction for this aurora. The centre of the view corresponds to the zenith at Barnstaple (N51°05′, W004°04′), and the geographic north is to the top. The outermost circle corresponds to the horizon.

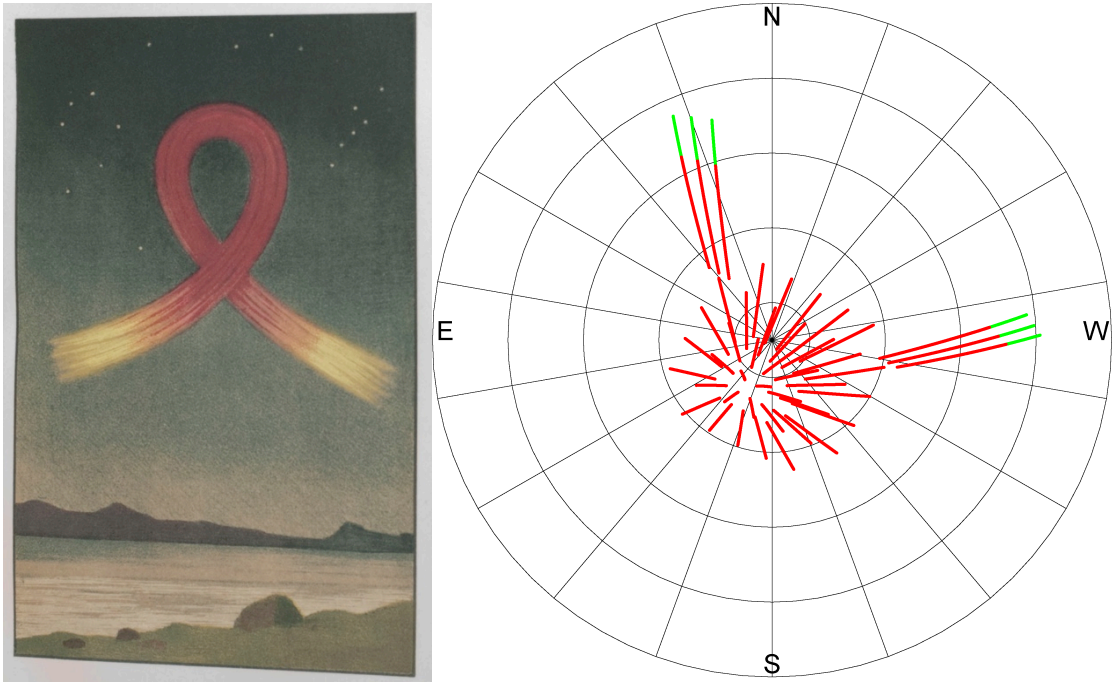

Figure 6: (left) The auroral drawing on 16 November 1871 by Lemström (1886); (right) Our reconstruction for this aurora with a shape of TP (tau-rho) or XP (chi-rho). The centre of the view corresponds to the zenith of Inari (N69°54′, E027°00′), and the geographic north is to the top. The outermost circle corresponds to the horizon.

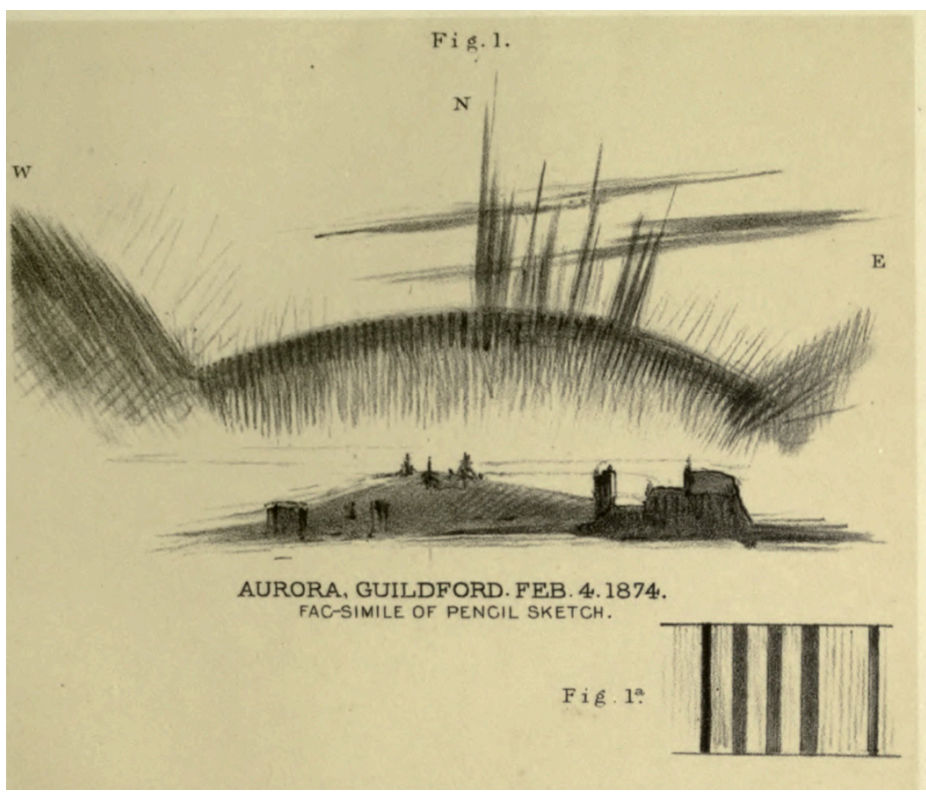

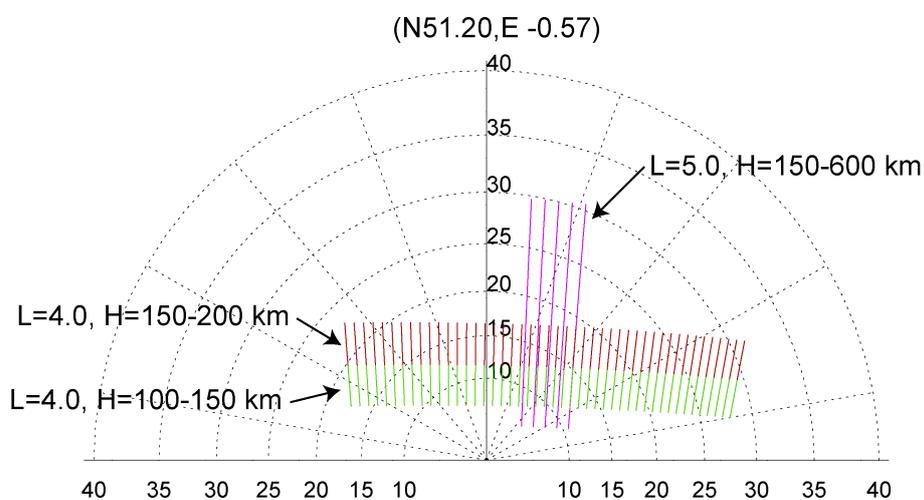

Figure 7: (above) The pencil sketch of the auroral display on 4 February 1874 by Capron (1879); (below) Our reconstruction of this cross-shaped aurora. The numerical figures represent the angle in degrees from the centre of the view. The centre of the view is placed 15° west from the geographical north. The observer is assumed to be located in Guildford (N51°12′, W000°36′).

**5.3. Auroral Displays during Twilight: Insights from Eyewitness Reports**

It is also worth considering if we can indeed reject auroral candidates during twilight, as claimed in N15b. Their discussion is probably based on one of the criteria for likeliness of an aurora in N15a (p.

230), which is "night-time observation (darkness)" stating "if the event clearly happened during civil or nautical twilight, we cannot conclude that it was an aurora".

However, early modern auroral reports indicate that auroral displays can be seen after sunset (Bottomley, 1872, p. 326), or during twilight with the unaided eyes, if they are bright enough. The first example is taken from a great magnetic storm on 1859 August 28, preceding the Carrington storm on 1/2 September 1859 (*e.g.*, Green and Boardsen, 2006; Green *et al*., 2006; Humble, 2006; Nevanlinna, 2008). During this storm at Beechworth (S36°22′, E146°52′; −45.7° MLAT), it was recorded that "Aug. 29th. Aurora visible for nearly an hour and a half, commencing about 5h 45m P. M., gradually increasing in beauty and brilliancy of tint until shortly before 7h, when the rays became gradually indistinct, disappearing at about 7h 15m P. M.. During the whole day the telegraph wires were strongly affected" (Loomis, 1861, p. 80). As shown in Table 4, this report indicates that an aurora was visible during civil twilight.

The same storm was described as follows: "Aug. 28th, the aurora was visible in the evening twilight especially to N. and N.E." at Gettysburg (N39° 49′, W110° 15′; 50.9° MLAT) by the Rev. M. Jacobs (Loomis, 1860, p. 345), and "While the evening twilight was yet so strong as to make the phenomenon scarcely discernible, a rosy hue was seen spreading over a space reaching from the northeastern horizon to the north star and thence to my zenith, of uniform breadth throughout, and bounded south by a line through Alpha Lyrae, passing vertically down to the east" at West Point of New York (N41°23′ W73°57′; 52.6° MLAT) by Prof. Alexander Twining (Loomis, 1859, p. 394).

On 4 February 1872, the aurora had been already observed since 17:00 LT at Blencowe (N54°41′, W002°51′; 54.7° MLAT), Hartlepool (N54°41′, W001°13′; 54.7° MLAT), and Monach (N57°31′, W007°34′; 57.5° MLAT). These observations are reported as: "At five o'clock a muddy undefined redness made its appearance in the N. E. and W., especially in the former, which continued for some time without any very marked change. Towards half-past six the redness became more concentrated, gradually brightened, and finally became of a most intense brilliancy" at Blencowe (Fawsett, 1872, p. 302), "But still earlier was it observed at Hartlepool, whence a correspondent writes, at 5 o'clock: — 'The whole of the southern sky was tinged with a most beautiful rose colour, which, as darkness set in, extended towards the zenith, where it culminated in a brilliant corona'" at Hartlepool (Earwaker, 1872, p. 322), and "at Monach, the most western island of the Hebrides, and at nearly all the 150 stations which report to the Society, appearing at some places as early as 5 p.m., and continuing visible at others till half-past one in the morning of the 5th" (Buchan, 1872, p. 461) respectively.

During the extreme magnetic storm on 25 October 1870, Fuertes Acevedo, a Spanish physicist at Santander, reported "brilliant skylights toward the north of a reddish violet color" at 17:40 LT before the end of twilight as well (Vaquero *et al.*, 2008, p. 7). The auroral display was reported at Lisbon. Luiz (1870, p. 129) at the Lisbon Observatory described it as "first clear at 17:45 [LT], up to the height of 60°".

Each observation is accompanied by great magnetic disturbances and simultaneous auroral observations and is hence securely considered as a genuine auroral display (Cliver and Dietrich, 2013; Silverman, 2008; Vaquero *et al.*, 2008; Hayakawa *et al.*, 2016a, 2018d, 2018f). Table 4 shows the calculated timing of sunset and the ends of civil twilight, nautical twilight, and astronomical twilight. We assume the altitude of the solar centre is -0.8° for sunset, allowing about 0.55° for refraction, and 0.25° for the semi-diameter. Likewise, we assume -6° for the end of civil twilight, -12° for the end of nautical twilight, and -18° for the end of astronomical twilight. We do not consider the difference between terrestrial time (TT) and universal time (UT) ($\Delta T = TT - UT$; see Stephenson, 1997) because it is less than 1 minute in the late 19th century (Morrison and Stephenson, 2004; Stephenson, Morrison, and Hohenkerk, 2016).

The auroral record at Beechworth on 29 August 1859 shows that an aurora was visible throughout the duration of civil twilight, while other observations for this storm do not specify the timing explicitly. Likewise, the onset of auroral observations at Blencowe, Hartlepool, and Monach on 4 February 1872 fell into the interval of civil twilight. The start of auroral observations at Santander and Lisbon on 25 October 1870 fell into the interval of nautical twilight. The visibility of the auroral displays during twilight is also confirmed instrumentally with modern all-sky auroral observations at South Pole Station (90° S) as detailed in Appendix 6 (see Ebihara *et al.*, 2007).

| Year | Month | Date | Place | Lat. | Long. | Sunset | CT | NT | AT |
|---|---|---|---|---|---|---|---|---|---|
| 1859 | 8 | 29 | Beechworth | S36°22′ | E146°52′ | 17:37 | 18:03 | 18:33 | 19:03 |
| 1870 | 10 | 25 | Santander | N43°28′ | W003°49′ | 17:02 | 17:31 | 18:05 | 18:38 |
| 1870 | 10 | 25 | Lisbon | N38°43′ | W009°08′ | 17:09 | 17:36 | 18:07 | 18:38 |
| 1872 | 2 | 4 | Blencowe | N54°41′ | W002°51′ | 16:44 | 17:23 | 18:06 | 18:48 |
| 1872 | 2 | 4 | Hartlepool | N54°41′ | W001°13′ | 16:44 | 17:23 | 18:06 | 18:48 |
| 1872 | 2 | 4 | Monach | N57°31′ | W007°34′ | 16:33 | 17:15 | 18:02 | 18:47 |

Table 4: The timing of sunset and ends of CT (civil twilight), NT (nautical twilight), and AT (astronomical twilight). The time shown in this table is all based on local mean time (*c.f.*, Silverman, 1998; Humble, 2006). These calculated values were cross-checked by the code by Nagasawa (2000) as well.

**6. Another Celestial Sign in the *Annales Regni Francorum* in 776**

The discussions in our paper cast a caveat on the discussions of another celestial sign in 776. N15a and N15b criticized U13's interpretation of the "manifestation of God's glory" namely "the likeness of two shields red with flame wheeling over the church" in the *Annales Regni Francorum* as auroral display (here after ARF; Pertz and Kurz, 1895, pp. 44-46; Scholz and Rogers 1972, pp. 53-55) as "Obviously, this happened during the day (as noticed by Gibbons & Werner 2012), as mentioned explicitly (on a day or quadam die)" and hence this cannot be auroral display due to its sky brightness (N15a, p. 240).

However, when we consult the translation by Scholz and Rogers (1972, p.53), this term *quadam die* (derived from *quidam dies*) is translated not as "during the day" but as "one day", which does not restrict the observational time to either daytime or night-time. Actually, the first word *quadam* is the singular feminine ablative of *quidam*, which has the meaning of "a certain/particular", which is combined with time, place, and person without the meaning of "during" (Lewis and Short, 1879, p. 1511; Glare, 1982, pp. 1551-1552). The second word *die* is singular feminine ablative of *dies*, which certainly means "day", while this word may mean both "the civil day of twenty-four hours" and "A natural day, a day, as opp. to night" (Lewis and Short, 1879, pp. 573-574; c.f. Glare, 1982, p. 539), just like "day" in English. Therefore, we consider the criticism to U13 by N15a and N15b to be illogical, based on the trimming of the meaning of Latin words.

As the record is fragmentary, we refrain from guessing about the interpretation of this celestial sign. However, at least we cannot exclude the possibility of night-time observations, if relying upon the original Latin text. It is not clear whether it was before sunset or after sunset in this case when the Saxons "prepared for battle against the Christians in the castle" (Scholz and Rogers, 1972, p. 53), as there is a mention to the battles in the night time concerned on the German tribes according to the Byzantine strategic accounts and show us that the time of battle is not necessarily confined to daytime (*e.g.*, Dennis, 1981, pp. 308-311, pp. 338-341, and p. 371; Dennis, 1984, p. 95, p. 107, and p. 119). In ARF, the celestial sign was described as a manifestation of God's glory, while the shape is hardly a cross. This is consistent with our discussions on the shape of the "sign of Christ", where we find not only a vertical cross but also other forms in contemporary usage in Anglo-Saxon England too.

**7. Solar Activity around the 770s**

So far, we have examined the various reports of variants of the celestial sign in the 770s. While the interpretation of this celestial sign has been controversial, we have shown that the original text favours not the halo hypothesis but the auroral hypothesis, by comparison with the observational evidence of parallel records in early modern times.

On the other hand, we consider the time range of this celestial sign, between 25 March 775 and 25 December 777, to be later than the expected input for the anomalous concentration of radiocarbon in 774/775, which is considered as around boreal Summer in 774: July ±1 month in 774 in Büngten *et al.* (2018), around early September in 774 in Sukhodolov *et al.* (2017), or late spring to early summer in 774 in Uusitalo *et al.* (2018). Indeed, U13 (p. 3) did "not directly associate any particular aurora with the $^{14}$C event".

However, the records in the ASC favour "a distinct cluster of aurorae between AD770 and AD776" and "a high solar activity level around AD775" (U13, p. 3), rather than the interpretation of "no aurorae from AD 774 to 785" or a "Schwabe cycle activity minimum" around 774±1 by N15a (pp. 244-245). Other contemporary auroral reports such as those in 770/771, 771/772, and 773 in the Zūqnīn Chronicle from North Mesopotamia (Harrak, 1999; Hayakawa *et al.*, 2017b), and 776 in Chinese official histories (Stephenson, 2015; Stephenson *et al.*, 2019) support this conclusion.

At the same time, it is not extreme coronal mass ejections (CMEs) but extreme solar energetic particles (SEPs) that are expected to be responsible for the anomalous concentration of radiocarbon in 774/775 (Miyake, Masuda, and Nakamura, 2013; U13; Mekhaldi *et al.*, 2015; Büngten *et al.*, 2018). The relationship between CMEs and SEPs is not straightforward. Gopalswamy *et al.* (2012) surveyed the ground level enhancement (GLE) of SEP events and the intensity of CMEs during Solar Cycle 23. As a result of this survey, Gopalswamy *et al.* (2012) found moderate to poor correlation between GLE intensity and flare size and CME speed in their Figure 6 (see also, Vennerstrom *et al.* (2016)). Moreover, even if SEP events were followed by interplanetary coronal mass ejections (ICMEs), the interplanetary magnetic field (IMF) needs to be southward to generate a great geomagnetic disturbance (Daglis, 2004; Lockwood *et al.*, 2016). The fast ICME in August 1972 failed to generate a great geomagnetic disturbance due to the northward direction of the IMF, while it had the potential to cause a great magnetic storm comparable to the Carrington event (Tsurutani *et al.*, 2003; Knipp *et al.*, 2018). In fact, we should note that the Carrington event in 1859 has not shown significant footprint in the isotope data (*e.g.* Usoskin *et al.*, 2006; Usoskin and Kovaltsov, 2012; McCracken and Beer, 2015) and we need to be careful to conclude that the ICME during the Carrington event was accompanied by SEPs.

Therefore, we stress the enhanced level of solar activity around 774/775, although we do not relate any of the specific auroral reports with the cause of the anomalous concentration of radiocarbon in 774/775. Even if this anomalous concentration is caused by extreme SEP events with a hard spectrum, a geomagnetic storm is not necessarily generated, as the between the GLEs originated by a SEP event and the intensity of the ICME is far from straightforward (Gopalswamy *et al.*, 2012) and ICMEs need the IMF to be southward to generate great geomagnetic storms (Tsurutani *et al.*, 2003; Daglis, 2004; Lockwood *et al.*, 2016).

**8. Conclusion**

In this article, we have examined the reports of the celestial sign in the 770s in variants of the ASC. We have confirmed that it is described as a red sign of Christ and its shape is not necessarily restricted to be a vertical cross, but can be also represented a diagonal cross, a cross with wedges in angles, or other divine signs such as XP (chi-rho) or TP (tau-rho), based on contemporary images in coins and manuscripts. The observational time of the sign was described literally as after sunset and the philological evidence does not support the interpretation of during sunset in N15a and N15b. The time range is confirmed as around 776, namely between 25 March 775 and 25 December 777, by considering the various possibilities of contemporary calendar systems (Hampson, 1841).

These philological and iconographic evidences favour not the halo hypothesis but the auroral hypothesis. The solar-halo hypothesis is rejected because of the observational time. Even if we expect the "red sign of Christ" as a vertical cross "at least with horizontal arc and vertical pillar" as in N15a, the horizontal arc of the solar halo is already below the horizon and is not visible after sunset. The lunar halo hypothesis is also unlikely because of its colour.

On the other hand, these descriptions do not contradict an auroral interpretation. The auroral display on 4 February 1872 was described as a "beautiful Maltese cross" at Bodmin. The shapes like crosses, chi-rho, or tau-rho are confirmed in the auroral drawings of the 19th century and have been reconstructed by using an appropriate magnetic field model. Moreover, we have observational evidence of auroral visibility "after sunset" in an early-modern observation in 1872. The eyewitness reports in 19th century and modern instrumental observations confirm that auroras can be visible even during twilight, if they are bright enough. This casts serious doubts on the newly suggested "criteria" for likeliness of an aurora in N15a.

Likewise, we briefly revisited another celestial sign in the *Annales Regni Francorum* and confirmed that the criticism of N15a against U13 is not valid, if we indeed consult the relevant Latin dictionaries.

Nevertheless, we do not directly relate this celestial observation with the cosmic ray event in 774/775, as the possible time range of this celestial sign is later than the expected onset of the cosmic ray event in the boreal Summer of 774 (*e.g.* Sukhodolov *et al*., 2017; Büngten *et al*., 2018; Uusitalo *et al*., 2018). Rather the present report suggests enhanced solar activity around 774/775, together with other historical auroral reports as suggested by U13, unlike N15a who placed a solar minimum around 774. This is in a good agreement with the independent reconstruction from Be data in ice cores, placing the solar maximum around 776 -- 777 as shown in Figure 1 of Sukhodolov et al. (2017). Notably, Büngten *et al*. (2018) consider that this is when the contemporary astronomical interests are enhanced, as well.

Finally, we must stress that absence of evidence is not evidence of absence due to the fragmentary nature of historical records in the 8th century (*e.g.* Silverman, 1992; S15; Barnard *et al*., 2017). Moreover, SEP events are not necessarily followed by great auroral displays, as the relationship between GLEs coming from SEP events and magnetic storms caused by CMEs with southward IMF is far from straightforward (*e.g*., Gopalswamy *et al*., 2012). Indeed, further historical researches around 774/775 in comparison with actual observational evidence during the known extreme space weather events would be beneficial to obtain better understanding of the level of the solar activity that far back in time.

**Acknowledgement**

We thank Corpus Christi College (the University of Cambridge), the British Library, the Fitzwilliam Museum (the University of Cambridge), and Staatsbibliothek zu Berlin (SBB) for providing the images and the permissions of researches of the manuscripts of Anglo-Saxon Chronicle, those of Anglo-Saxon coins, and those of *Annales Laudensis et Sancti Vincentii*. We gratefully acknowledge L. Cowley for providing HaloSim 3, R. Naismith for his advice on the contemporary coinage, A. Popescu for permission of photographing Anglo-Saxon coins in the Fitzwilliam Museum (the University of Cambridge), the archivist in SBB for his permission of reproduction and advices on precise referencing of the MS Phillipps 1830, Y. Okada for his helps on accessing German documents, S. Kikuchi for his advices on the Carolingian military systems, the Carolingian manuscripts, and medieval Latin dictionaries for the Carolingian territories, K. Nakada for his advice on the medieval European military systems, and T. Hoshi for providing calculation code for sunset and sunrise. This work was supported by the UKSSDC, a NSF grant ANT-1643700, and a Grant-in-Aid from the Ministry of Education, Culture, Sports, Science and Technology of Japan, Grant Number JP18H01254 (PI: H. Isobe), JP15H05816 (PI: S. Yoden), JP15H03732 (PI: Y. Ebihara), JP16H03955 (PI: K. Shibata), and JP15H05815 (PI: Y. Miyoshi), and a Grant-in-Aid for JSPS Research Fellows JP17J06954 (PI: H. Hayakawa).

**Disclosure of Potential Conflict of Interest**

The authors declare that they have no conflict of interest.

**References**

Allen, J. 2012, *Nature* **486**, 473. doi:10.1038/nature.2012.10898

Baker, P. S., ed. 2000, *The Anglo-Saxon Chronicle 8 MS F: a semi-diplomatic edition with introduction and indices*. (Cambridge: Cambridge University Press)

Barnard L., Owens, M., Scott, C. J.: 2017, *Astronomy & Geophysics*, **58**, 2, 12, doi: 10.1093/astrogeo/atx056

Bately, J., ed. 1986, *The Anglo-Saxon Chronicle 3 MS A: a semi-diplomatic edition with introduction and indices*. (Cambridge: Cambridge University Press)

Bernhardt, J. W. 1993, Itinerant Kingship and Royal Monasteries in Early Medieval Germany, c. 936–1075, *CSMLT 4th ser.*, **21**, 45.

Blackburn, M.: 2007, 'Crosses and Conversion: The Iconography of the Coinage of Viking York ca. 900', in Jolly, Karkov, and Keefer (2007, 172-200).

Bosworth, J., Toller, T. N.: 1921, *An Anglo-Saxon Dictionary: Based on the Manuscript Collections of the Late Joseph Bosworth: Supplement* (Oxford: Clarendon Press).

Bottomley, J. T. 1872, *Nature*, **5**, 326.

Bremmer Jr., R. H.: 2010, 'Old English "Cross" Words''', in Keefer, Jolly, and Karkov 2010, 204-232.

Brown, G. H.: 2007, 'Bede and the Cross', in Jolly, Karkov, and Keefer 2007, 19.

Brown, M. P.: 2010, 'The Cross and the Book: the Cross-Carpet Pages of the Lindisfarne Gospels as Sacred Figurae', in Keefer, Jolly, and Karkov (2010, 17).

Buchan, A.: 1872, *Nature*, **5**, 461.

Büngten, U., Wacker, L., Galvan, D. J., Arnold, S., Arseneault, D., Baillie, M., *et al*.: 2018, *Nature Communications*, **9**, 3605. doi: 10.1038/s41467-018-06036-0

Campbell, A., ed. and trans.: 1962, *The Chronicle of Æthelweard* (London: Nelson).

Capron, J. R.: 1879, *Aurorae, their characteristics and spectra* (London: E. & F.N. Spon).

Carrasco, V. M. S., Álvarez, J. V., Vaquero, J. M.: 2015, *Solar Physics*, **290**, 10, 2719. doi: 10.1007/s11207-015-0767-z

Carrasco, V. M. S., Vaquero, J. M., Gallego, M. C.: 2018, *Solar Physics*, **293**, 3, 51. doi: 10.1007/s11207-018-1270-0

Clark, D. H., Stephenson, F. R.: 1977, *The Historical Supernovae* (Oxford: Pergamon Press)

Clark Hall, J. R.: 1960, *A Concise Anglo-Saxon Dictionary*, 4th edn, with a supplement by H. D. Merrit (Cambridge: Cambridge University Press).

Clette, F., Svalgaard, L., Vaquero, J. M., Cliver, E. W.: 2014, *Space Science Reviews*, **186**, 1-4, 35. doi: 10.1007/s11214-014-0074-2

Cliver, E. W., Dietrich, W. F.: 2013, *J. Space Weather Space Clim.*, **3**, A31. doi: 10.1051/swsc/2013053

Cliver, E. W., Tylka, A. J., Dietrich, W. F., Ling, A. G.: 2014, *The Astrophysical Journal*, **781**, 1, 32. doi: 10.1088/0004-637X/781/1/32

Cubbin, G. P., ed.: 1996, *The Anglo-Saxon Chronicle 6 MS D: a semi-diplomatic edition with introduction and indices*. (Cambridge: Cambridge University Press)

Daglis, I. A.: 2004, *Effects of space Weather on Technology Infrastructure* (Amsterdam, Kluwer Academic Press).

Dammery, R. J. E.: 1991, *Law-code of King Alfred the Great (Doctoral thesis)*. doi: 10.17863/CAM.15910

Darlington, R. R., P. McGurk and J. Bray, eds. and trans.: 1995, *The Chronicle of John of Worcester, Volume II, The Annals from 450 to 1066* (Oxford: Oxford University Press).

Dee, M., Pope, B., Miles, D., Manning, S., Miyake, F.: 2017, *Radiocarbon*, **59**, 2, 293. doi: 10.1017/RDC.2016.50

Dennis, G. T., ed.: 1981, *Das Strategikon des Maurikios* (trans. E. Gamillschneg), Vienna: VÖAW.

Dennis, G. T.: 1984, *Maurice's Strategikon: Handbook of Byzantine Military Strategy*, Philadelphia: University of Pennsylvania Press.

Earwaker, J. P.: 1872, *Nature*, **5**, 322.

Ebihara, Y., Tanaka, Y. -M., Takasaki, S., Weatherwax, A. T., Taguchi, M.: 2007, *Journal of Geophysical Research*, **112**, A01201, doi: 10.1029/2006JA012087.

Ebihara, Y., Hayakawa, H., Iwahashi, K., Tamazawa, H., Kawamura, A. D., Isobe, H.: 2017, *Space Weather*, **15**, 1373. doi: 10.1002/2017SW001693

Enoto, T., Kisaka, A., Shibata, K.: 2018, *Reports on Progress in Physics*, under review.

Fawsett, T.: 1872, *Nature*, **5**, 302.

Fogtmann-Schulz, A., Østbø, S. M., Nielsen, S. G. B., Olsen, J., Karoff, C., Knudsen, M. F.: 2017, *Geophysical Research Letters*, **44**, 16, 8621. doi: 10.1002/2017GL074208

Freeborn, D. 1998, *From Old English to Standard English: A Course Book in Language Variation Across Time* (Ottawa: University of Ottawa Press)

Fulk, R. D.: 2009, *Florilegium*, **26**, 15.

Garmonsway, G. N. ed. & tr.: 1953, *The Anglo-Saxon chronicle* (London: Dent & Sons)

Gibbons G. W., Werner, M. C.: 2012, *Nature*, **487**, 432. doi: 10.1038/487432c

Glare, P. G. W.: 1982, *Oxford Latin Dictionary* (Oxford: Clarendon Press).

Gopalswamy, N., Xie, H., Yashiro, S., Akiyama, S., MäkeläIa, P., Usoskin, I. G.: 2012, *Space Sci. Rev.*, 171, 23. doi: 10.1007/s11214-012-9890-4

Green, J. L., Boardsen, S., Odenwald, S., Humble, J., Pazamickas, K. A.: 2006, *Adv. Space Res.*, **38**, 2, 145. doi: 10.1016/j.asr.2005.12.021

Green, J. L., Boardsen, S.: 2006, *Adv. Space Res.*, **38**, 130. doi: 10.1016/j.asr.2005.08.054

Greenaway, D., ed. and trans. 1996, *Henry Archdeacon of Huntingdon: Historia Anglorum* (Oxford: Oxford University Press).

Hall, T. M.: 1872, *Meteorological Magazine*, **7**, 73, 1.

Hambaryan, V. V., Neuhäuser, R.: 2013, *Monthly Notices of the Royal Astronomical Society*, **430**, 1, 32. doi: 10.1093/mnras/sts378

Hampson, R.T.: 1841, *Medii Aevi Kalendarium*, II (London: Henry Kent Caston and Co.).

Hayakawa, H., Mitsuma, Y., Fujiwara, Y., *et al*.: 2017b, *Publications of the Astronomical Society of Japan*, **69**, 17. doi: 10.1093/ pasj/psw128

Hayakawa, H., Iwahashi, K., Tamazawa, H., *et al*.: 2016a, *Publications of the Astronomical Society of Japan*, **68**, 99. doi: 10.1093/pasj/psw097

Hayakawa, H., Mitsuma, Y., Ebihara, Y., *et al*.: 2016b, *Earth, Planets and Space*, **68**, 195. doi: 10.1186/s40623-016-0571-5

Hayakawa, H., Ebihara, Y., Willis, D. M., *et al*.: 2018d, *The Astrophysical Journal*, **862**, 15. doi: 10.3847/1538-4357/aaca40

Hayakawa, H., Ebihara, Y., Hand, D. P., *et al*.: 2018f, *The Astrophysical Journal*, **869**, 57. doi: 10.3847/1538-4357/aae47c

Herbst, K. D.: 2006, *Die Korrespondenz des Astronomen und Kalendermachers Gottfried Kirch (1639–1710)*, Briefe 1689– 1709 (Garamond: Jena).

Hinton, D. A. 2010, 'The Anglo-Saxon Chapel at Bradford-on-Avon, Wiltshire', in (Keefer, Jolly, and Karkov, 2010, 319).

Howlett, D. R., Christschev, T., Evans, T., Piper, P. O., White, C.: 2009, *Dictionary of the Medieval Latin from British Sources*, **12** (Oxford: Oxford University Press).

Humble, J. E.: 2006, *Advances in Space Research*, **38**, 155. doi: 10.1016/j.asr.2005.08.053

Hunten, D. M.: 2003, *Planetary and Space Science*, **51**, 13, 887. doi:

10.1016/S0032-0633(03)00079-5

Irvine, S., ed.: 2004, *The Anglo-Saxon Chronicle 7 MS E: a semi-diplomatic edition with introduction and indices*. (Cambridge: Cambridge University Press)

Jackson, A., Jonkers, A. R. T., Walker, M. R.: 2000, *Roy. Soc. of London Phil. Tr. A*, **358**, 1768, 957. doi: 10.1098/rsta.2000.0569

Jolly, K. L., Karkov, C. E., Keefer, S. L., eds.: 2007, *Cross and Culture in Anglo-Saxon England: Studies in Honor of George Hardin Brown*, (Morgantown: West Virginia University Press).

Jull, A. J. T., Panyushkina, I. P., Lange, T. E., *et al.*: 2014, *Geophysical Research Letters*, **41**, 8, 3004. doi: 10.1002/2014GL059874

Karkov, C. E., Keefer, S. L., Jolly, K. L., eds.: 2006, *The Place of the Cross in Anglo-Saxon England* (Woodbridge: The Boydell Press).

Karoff, C., Knudsen, M. F., De Cat, P., Bonanno, A., Fogtmann-Schulz, A., Fu, J., *et al.*: 2016, *Nature Communications*, **7**, 11058. doi: 10.1038/ncomms11058

Keefer, S. L., K. L. Jolly, and C. E. Karkov, eds.: 2010, *Cross and Cruciform in the Anglo-Saxon England: Studies to Honor the Memory of Timothy Reuter* (Morgantown: Arizona University Press).

Knipp, D. J., Fraser, B. J., Shea, M. A., Smart, D. F.: 2018, *Space Weather*, **16**. doi: 10.1029/2018SW002024

Korte, M., Constable, C.: 2011, *Physics of the Earth and Planetary Interiors*, **188**, 3, 247. doi: 10.1016/j.pepi.2011.06.017

Kronk, G. W.: 1999, *Cometography*, **1** (Cambridge: Cambridge University Press).

Kurze, F.: 1895, *Annales regni Francorum inde ab a. 741 usque ad a. 829, qui dicuntur Annales Laurissenses maiores et Einhardi* in Scriptores rerum Germanicarum *in usum scholarum* 6, Monumenta Germaniae Historica (Hannover).

Lapidge, M., Blair, J., Keynes, S., Scragg, D., eds.: 2014, *The Wiley Blackwell Encyclopedia of Anglo-Saxon England, 2nd edn.* (Oxford: Willey Blackwell), 350, *s.v.* OLD ENGLISH

Lemström, S.: 1886, *L'aurore boréale: Étude générate des phénomènes produits par les courants électriques de l'atmosphère* (Paris: Gauthier-Villars).

Lenker, U.: 2010, 'Signifying Christ in Anglo-Saxon England: Old English Terms for the Sign of the Cross', in Keefer, Jolly, and Karkov (2010, 233).

Lewis, C. T., Short, C. 1879, *A Latin Dictionary*. (Oxford: Clarendon Press).

Link, F.: 1962, *Geofysikální Sborník*, **173**, 297

Liu, Y., Zhang, Z. F., Peng, Z. C., *et al.*: 2014, *Scientific Reports*, **4**, 3728. doi: 10.1038/srep03728

Lockwood, M., Owens, M. J., Barnard, L. A., Bentley, S., Scott, C. J. Watt, C. E.: 2016, *Space

*Weather*, **14**, 6, 406. doi: 10.1002/2016SW001375

Longmans and Roberts, 1858: *Rerum Britannicarum Medii Aevi Scriptores* (London: Longmans and Roberts).

Loomis, E.: 1859, *American Journal of Science, Second Series*, **28**, 84, 385-408.

Loomis, E.: 1860, *American Journal of Science, Second Series*, **30**, 90, 339-361.

Loomis, E.: 1861, *American Journal of Science, Second Series*, **32**, 94, 71-84.

Luiz, D. 1870, *Annaes do Observatorio do Infante D. Luiz*, v.8.

Maehara, H., Shibayama, T., Notsu, S., Notsu, Y., Nagao, T., Kusaba, S., *et al*.: 2012, Nature, **485**, 478. doi: 10.1038/nature11063

Maehara, H., Notsu, Y., Notsu, S., Namekata, K., Honda, S., Ishii, T. T., Nogami, D., Shibata, K.: 2017, *Publications of the Astronomical Society of Japan*, **69**, 41. doi: 10.1093/pasj/psx013

Maehara, H., Shibayama, T., Notsu, Y., Notsu, S., Honda, S., Nogami, D., Shibata, K.: 2015, *Earth, Planets, and Space*, **67**, 59. doi: 10.1186/s40623-015-0217-z

Marsden, R.: 2004, *The Cambridge Old English Reader* (Cambridge: Cambridge University Press).

McCracken, K. G., Beer, J.: 2015, *Solar Physics*, **290**, 10, 3051. doi: 10.1007/s11207-015-0777-x

Mekhaldi, F., Muscheler, R., Adolphi, F., Aldahan, A., Beer, J., McConnell, J. R., *et al*.: 2015, *Nature Communications*, **6**, 8611. doi: 10.1038/ncomms9611

Minnaert, M. G. J.: 1993, *Light and Color in the Outdoors* (New York: Springer).

Miyake, F., Jull, A. J. T., Panyushkina, I. P., Wacker, L., Salzer, M., Baisan, C. H., *et al*.: 2017, *Proceedings of the National Academy of Science*, **114**, 881. doi: 10.1073/pnas.1613144114

Miyake, F., Masuda, K., Nakamura, T.: 2013, *Nature Communications*, **4**, 1748. doi: 10.1038/ncomms2783

Miyake, F., Nagaya, K., Masuda, K., Nakamura, T.: 2012, *Nature*, **486**, 240. doi: 10.1038/nature11123

Miyake, F., Suzuki, A., Masuda, K., Horiuchi, K., Motoyama, H., Matsuzaki, H., *et al*.: 2015, *Geophysical Research Letters*, **42**, 1, 84. doi: 10.1002/2014GL062218

MMM Editorial: 1872, *Monthly Meteorological Magazine*, **1872**, 32-35.

Morrison, L. V., Stephenson, F. R.: 2004, *Journal for the History of Astronomy*, **35**, 3, 120, 327 – 336. doi: 10.1177/002182860403500305

Nagasawa, K. 2000: *Calculation of Sunset and Sunrise* (Tokyo: Chijin Shokan).

Naismith, R. ed.: 2017, *Medieval European Coinage with a Catalogue of the Coins in the Fitzwilliam Museum, Cambridge: v.8 Britain and Ireland c. 400–1066* (Cambridge: Cambridge University Press).

Namekata, K., Sakaue, T., Watanabe, K., Asai, A., Maehara, H., Notsu, Y., *et al*.: 2017, *The Astrophysical Journal*, **851**, 91. doi: 10.3847/1538-4357/aa9b34

Neuhäuser, R., Neuhäuser, D. L. 2015a, *Astronomische Nachrichten*, **336**, 3, 225. doi: 10.1002/asna.201412160 (N15a)

Neuhäuser, D. L., Neuhäuser, R. 2015b, *Astronomische Nachrichten*, **336**, 10, 913. doi: 10.1002/asna.201512205 (N15b)

Nevanlinna, H. 2008, *Adv. Space Res.*, **42**, 171. doi: 10.1016/j.asr.2005.07.076

Niermeyer, J. F. 1976, *Mediae Latinitatis lexicon minus* (Leiden, Brill).

Notsu, Y., Honda, S., Maehara, H., Notsu, S., Shibayama, T., Nogami, D., Shibata, K.: 2015a, *Publications of the Astronomical Society of Japan*, **67**, 3, 32, doi: 10.1093/pasj/psv001

Notsu, Y., Honda, S., Maehara, H., Notsu, S., Shibayama, T., Nogami, D., Shibata, K.: 2015b, *Publications of the Astronomical Society of Japan*, **67**, 3, 33, doi: 10.1093/pasj/psv002

Otto, A., Lummerzheim, D., Zhu, H., Lie-Svendsen, Ø., Rees, M. H., Lanchester, B. S.: 2003, *J. Geophys. Res.*, 108, 8017, A4. doi: 10.1029/2002JA009423.

Owen-Crocker, G. R. Stephens, W.: 2007, 'The Cross in the Grave: Design or Divine?,' in: Jolly, Karkov, and Keefer (2007, 118).

Owens, B. 2013, *Nature*, **495**, 7441, 300. doi:10.1038/495300a

O'Brien O'Keeffe, K., ed.: 2001, *The Anglo-Saxon Chronicle 5 MS C: a semi-diplomatic edition with introduction and indices*. (Cambridge: Cambridge University Press)

Pavlov, A.K., Blinov, A.V., Konstantinov, A.N., *et al.*: 2013, *Mon. Not. Roy. Astron. Soc.* **435**, 2878. doi: 10.1093/mnras/stt1468.a

Pertz, G. H.: 1841, *Monumenta Germaniae Historica*, SS IV (Hannoverae: Impensis Bibliopolii Hahniani)

Pertz, G. H.: 1888, *Monumenta Germaniae Historica*, SS XV (Hannoverae: Impensis Bibliopolii Hahniani)

Pertz, G. H., Kurz, F.: 1895, *Monumenta Germaniae Historica*, SS rer. Germ. VI (Hannoverae: Impensis Bibliopolii Hahniani)

Peyer, H. C.: 1964, 'Das Reisekönigtum des Mittelalters', *Vierteljahrschrift für Sozial-und Wirtschafsgeschichte*, **51**, 1.

Roach, L.: 2013, *Kingship and Consent in Anglo-Saxon England, 871-978* (Cambridge, Cambridge University Press), 45.

Rodríguez, D. M.: 2018, *Lexicography ASIALEX*, **4**, 119. doi: 10.1007/s40607-018-0043-0

Rumble, A. R.: 2006, 'The Cross in English Place-names: Vocabulary and Usage', in Karkov,

Keefer, and Jolly (2006, 29).

Sawyer, P.: 1968, *Anglo-Saxon Charters: an Annotated List and Bibliography* (London: Royal Historical Society).

Scholz, B., trans. with Rogers, B.: 1972, *Carolingian chronicles: Royal Frankish annals and Nithard's Histories* (Ann Arbor: University of Michigan Press).

Shaw, P. A.: 2013, *Early Medieval Europe*, **21**, 2, 115. doi: 10.1111/emed.12012

Shibata, K., Isobe, H., Hillier, A., S., Choudhuri, A. R., Maehara, H., Ishii, T. T., *et al*.: 2013, *Publications of the Astronomical Society of Japan*, **65**, 49. doi: 10.1093/pasj/65.3.49

Sigl, M., Winstrup, M., McConnell, J. R., Welten, K. C., Plunkett, G., Ludlow, F., *et al*.: 2015, *Nature*, **523**, 7562, 543. doi: 10.1038/nature14565

Silverman, S. M.: 1992, *Reviews of Geophysics*, **30**, 4, 333. doi: 10.1029/92RG01571

Silverman, S. M.: 1998, *Journal of Atmospheric and Solar-Terrestrial Physics*, **60**, 10, 997. doi: 10.1016/S1364-6826(98)00040-6

Silverman, S. M., Cliver, E. W.: 2001, *Journal of Atmospheric and Solar-Terrestrial Physics*, **63**, 523. doi: 10.1016/S1364-6826(00)00174-7

Silverman, S. M.: 2006, *Advances in Space Research*, **38**, 2, 200. doi: 10.1016/j.asr.2005.03.158

Silverman, S. M.: 2008, *Journal of Atmospheric and Solar-Terrestrial Physics*, **70**, 10, 1301. doi: 10.1016/j.jastp.2008.03.012

Smart, D. F., Shea, M. A., McCracken, K. G.: 2006, *Advances in Space Researches*, **38**, 2, 215. doi: 10.1016/j.asr.2005.04.116

Stephenson, F. R., Green, D. A.: 2002, *Historical supernovae and their remnants* (Oxford, Oxford University Press)

Stephenson, F. R., Willis, D. M. and Hallinan, T. J.: 2004, *Astronomy & Geophysics*, **45**, 6.15. doi: 10.1046/j.1468-4004.2003.45615.x

Stephenson, F.R.: 2015, *Adv. Space Res.*, **55**, 1537. doi: 10.1016/j.asr.2014.12.014. (S15)

Stephenson, F.R., Morrison, L.V., Hohenkerk, C.Y.: 2016, *Proc. R. Soc. A*, **472**, 20160404; doi: 10.1098/rspa.2016.0404

Stephenson, F. R., Willis, D. M., Hayakawa, H., Ebihara, Y., Scott, C. J., Wilkinson, J., Wild, M. N.: 2019, *Solar Physics*, doi: 10.1007/s11207-019-1417-7.

Stubbs, W., ed.: 1868, *Chronica magistri rogeri de Houdene*, v.1 (London; Longmans etc.).

Størmer, C.: 1955, *The Polar Aurora*. (Oxford, Oxford University Press)

Sukhodolov, T., Usoskin, I. G., Rozanov, E., Asvestari, E., Ball, W. T., Curran, M. A. J., *et al*.: 2017, *Scientific Reports*, **7**, 45257. doi: 10.1038/srep45257

Swanton, M. J.: 2000, *The Anglo-Saxon Chronicles* (London: Phoenix).

Tape, W.: 1994, *Atmospheric Halos, Antarctic Research Series, Vol. 64*, (Washington DC: American Geophysical Union).

Taylor, S., ed.: 1983, *The Anglo-Saxon Chronicle 4 MS B: a semi-diplomatic edition with introduction and indices*. (Cambridge: Cambridge University Press)

Thomas B. C., Melott A. L., Arkenberg K. R. Snyder B. R.: 2013, *Geophys. Res. Lett.* **40**, 1237. doi: 10.1002/grl.50222

Treschow, M.: 1994, *Florilegium*, **13**, 79.

Tsurutani, B. T., Gonzalez, W. D., Lakhina, G. S., Alex, S. 2003 *J. Geophys. Res.*, **108**, 1268, doi:10.1029/2002JA009504.

Usoskin, I. G., Solanki, S. K., Kovaltsov, G. A., Beer, J., Kromer, B.: 2006, *Geophysical Research Letters*, **33**, 8, L08107. doi: 10.1029/2006GL026059

Usoskin, I. G., Arlt, R., Asvestari, E., et al. 2015, *Astronomy & Astrophysics*, **581**, A95. doi: 10.1051/0004-6361/201526652

Usoskin, I. G., Kovaltsov, G. A., Mishina, L. N., Sokoloff, D. D., Vaquero, J. M.: 2017, *Solar Physics*, **292**, 1, 15. doi: 10.1007/s11207-016-1035-6

Usoskin, I. G., Kovaltsov, G. A. 2012, *The Astrophysical Journal*, **757**, 92. doi: 1088/0004-637X/757/1/92

Usoskin, I. G., Kromer, B., Ludlow, F., Beer, J., Friedrich, M., Kovaltsov, G. A., Solanki, S. K., Wacker, L.: 2013, *Astronomy & Astrophysics*, **552**, L3. doi: 10.1051/0004-6361/201321080 (U13)

Uusitalo, J., Arppe, L., Hackman, T., Helama, S., Kovaltsov, G., Mielikäinen, K., *et al*.: 2018, *Nature Communications*, **9**, 3495. Doi: 10.1038/s41467-018-05883-1

Vaquero, J. M.; 2007, *Advances in Space Research*, **40**, 7, 929. doi: 10.1016/j.asr.2007.01.087

Vaquero, J. M., Valente, M. A., Trigo, R. M., Ribeiro, P., Gallego, M. C.: 2008, *Journal of Geophysical Research: Space Physics*, **113**, A8, A08230. doi: 10.1029/2007JA012943

Vaquero, J. M., Vazquez, M.: 2009, *The Sun Recorded Through History* (Berlin: Springer).

Vaquero, J. M., Kovaltsov, G. A., Usoskin, I. G., Carrasco, V. M. S., Gallego, M. C.: 2015, *Astronomy & Astrophysics*, **577**, A71. doi: 10.1051/0004-6361/201525962

Vaquero, J. M., Svalgaard, L., Carrasco, V. M. S., *et al*.: 2016, *Solar Physics*, **291**, 9-10, 3061. doi: 10.1007/s11207-016-0982-2

Vennerstrom, S., Lefevre, L., Dumbović, M., *et al*.: 2016, *Solar Physics*, **291**, 5, 1447. doi: 10.1007/s11207-016-0897-y

Waitz, G., Kehr, K. E., Hirsch, P., Lohmann, E.-H.: 1935, *Monumenta Germaniae Historica*, SS rer.

Germ. LX (Hannoverae: Impensis Bibliopolii Hahniani)

Whitelock, D.: 1961, *The Anglo-Saxon Chronicle* (London: Jarrold & Sons Ltd).

Whitelock, D.: 1979, *English Historical Documents*, vol. 1, *c. 500-1042* (Oxford: Oxford University Press)

Willis, D.M., Stephenson, F.R.: 2001, *Ann. Geophys.*, **19**, 289. doi: 10.5194/angeo-19-289-2001

Willis, D. M., Easterbrook, M. G., Stephenson, F. R.: 1980, *Nature*, **287**, 5783, 617-619. doi: 10.1038/287617a0

Yau, K., Stephenson, F. R., Willis, D. M.: 1995, *A catalogue of auroral observations from China, Korea and Japan (193 B.C. - A.D. 1770)* (Rutherford Appleton Lab., Chilton), ISSN 1358-6254,

Zolotova, N. V., Ponyavin, D. I.: 2015, *The Astrophysical Journal*, **800**, 1, 42. doi: 10.1088/0004-637X/800/1/42

Zolotova, N. V., Ponyavin, D. I.: 2016, *Solar Physics*, **291**, 9-10, 2869. doi: 10.1007/s11207-016-0908-z

Zupitza, J., ed.: 1880, *Ælfric's Grammatik und Glossar, Sammlung englischer Denkkmäler in kritischer Ausgabe* (Berlin: Weidmann).

▪▪▪▪▪▪▪▪▪▪▪▪▪▪▪▪▪▪▪▪▪▪▪▪▪▪▪▪▪▪▪▪▪▪▪▪▪▪▪▪▪▪▪▪▪▪▪▪▪▪▪▪▪▪▪▪▪▪▪▪▪▪▪▪

**Appendix 1: Records of the Celestial Sign in the ASC and their Translation**

**Appendix 1.1.: Variants of the ASC**

MS A, *f.* 10v.: [773] *AN. .dcclxxiii. Her oþiewde read Cristesmęl on hefenum æfter sunnan setlgonge*. (v.3, p.39)

MS B, *f.* 12v.: <774> *Her oðeowde read Cristes mæl on heofonum æfter sunnansetlgange*. (v.4, p.27)

MS C, *f.* 126v.: [774] *AN. dclxxiiii. Her oðywde read Criste<s> mæl on heofonum æfter sunnan setlgange*. (v.5, p.49)

MS D, *f.* 25r.: [774] *AN. .dcclxiiii. ... <; 7 men gesegon> read Cristes mel on hoefenum æfter sunnan setlgange*. (v.6, p.15)

MS E, *f.* 24r.: [774] *AN. dcclxxiiii. 7 men gesegon read Cristes mel on heofenum æfter sunnan setlangange*. (v.7, p.39)

MS F, *f.* 48v.: [774] *.dcclxxiiii. ⋯ 7 me'n'n gesegan read Cristes mæl on heouonum æfter sunnan setlegange;// ... et uisum 'est' crucis signum in cęlo post solis occubitum;* (v.8, p.52)

**Appendix 1.2: Translations of Variants of the ASC**

MS A-C

773 [776] Here, a red sign of Christ appeared in the heavens after the sun's setting. (Swanton, 2000, p.50)

MS D-F: And [776] men saw a red sign of Christ in the heavens after the sun's setting. (Swanton, 2000, p.51)

MS F Latin: The sign of a cross was seen in the heaven after the sun's setting.

**Appendix 2: Records of the Celestial Sign in Associated Chronicles**

(1) Chronicle of Æthelweard, compiled in c. 975, (see Campbell, 1962, p. xiii; p.25)

*Transcription*: 772 ..., *in coelo signum dominicæ crucis post solis occasum, et ipso anno bellum ciuile gestum est inter Cantuarium populum et Merciorum in cognominato loco Ottanforda.*

*Translation*: 772 ... the sign of the Lord's cross appeared in the sky after sunset, and in the same year civil war was waged between the people of Kent and that of the Mercians in the place called Otford.

(2) The Chronicle of John of Worcester, compiled in c. 1140 (Darlington, McGurk, and Bray 1995, pp. 210-211)

*Transcription*: [774] (vi) *796 Rubicundi coloris signum in crucis modum in celo apparuit post solis occasum.*

*Translation*: A red sign after the fashion of a cross appeared in the sky after sunset.

(3) Henry of Huntingdon, *Historia Anglorum*, compiled in *c*. 1130-1154 (Greenaway 1996, pp. 250-251)

*Transcription*: … *Hoc autem anno uisa fuerant in celo rubea signa post occasum solis*…

*Translation*: … In this year, however, red signs had been seen in the sky after sunset…

**Appendix 3: The Variation of "the Sign of Christ" and "the Sign of the Cross"**

As shown in the section 4.1, "the sign of Christ" and "the sign of cross" are not restricted to the vertical cross. Lenker (2010, p. 263) enters a caveat on its interpretation: "interpretation of *cristes mæl* as the "cross" requires the ideas stressed by Ælfric[3] […] that the holy rood is a sign for Christ. Without this interpretative support in a theological background, *cristes mæl* could, for instance, also denote baptism (*cf.* German *Christusmal* "baptism" in theological discourse) or be taken to refer to the stigmata of Christ (*cf.* German *Wundmal*)". Indeed, in one of the Old English charms recorded in a medical text called Bald's Leechbook in a 10th-century manuscript, "[t]he author seems to distinguish in meaning between *Cristes mæl* and *cruc* [Old English word for a cross]" (Bremmer 2010, pp. 213-214). In this text, it is commanded that *Cristes mæl* should be made on every limb, which might symbolize the wounds Christ suffered on the cross (stigmata). In short, our "red sign of Christ" could also be his stigma, with blood flooding from it, hence red.

---

[3] Ælfric of Eynsham, one of the greatest scholars of the Anglo-Saxon period and famous for many prose works.

Meanwhile, the majority of the examples of *Cristes mæl* and its variant c*ristelmæl* attested in Anglo-Saxon sources did mean the sign of the cross and some scholars attribute this meaning to the celestial sign (Lenker 2010, p. 266, n. 91). Nevertheless, a cross itself can take various forms too. Etymologically speaking, Greek "*stauros* primarily means "stake, pole" and, by extension, any kind of elevated structure on which criminals were executed, including the time-honoured representation of the cross, but also, for example, the CHI-shaped cross X, on which the apostle Andrew allegedly suffered his death, and the TAU-shaped cross T" (Bremmer, 2010, p. 205). Also, "The Latin word *crux*, too, originally had little to do with our word "cross", but signified, like the Greek *stauros*, a "pole" or "stake"" (Bremmer, 2010, p. 206).

Anglo-Saxons recognized crosses with various shapes as well. The Old English word *rōd* again "originally signified 'pole' or 'stake'. As such, it semantically parallels Greek *stauros* and Latin *crux*, something that Ælfric, for example, was well aware of: in his *Glossary* he included "*crux oððe stáurus: rod*" (Bremmer, 2010, p.216; Zupitza, 1880, p.313). The Anglo-Saxon's flexible attitude towards "cross" is extant as well in their writings and objects. Both Bede, one of the best scholars in early medieval Europe, and Ælfric perceived the Greek letter *Tau* as a cross (Brown 2007, p. 27; Lenker, 2010, pp. 258-259, n. 59). An artist-scribe of the Lindisfarne Gospels rendered its carpet pages with various forms of the cross, *i.e.* the Latin cross, the Greek cross, the ring-headed or solar cross, and the tau cross, also known as the Egyptian cross or the cross of St. Anthony (Brown 2010, pp. 33-52, and Figs. 1.1-5). Alongside the usual cross, various cruciform monograms like chi-rho (☧), tau-rho (staurogram), IX (✳) were employed as designs on coins (Blackburn, 2007, pp. 185-199) and vessels (Owen-Crocker and Stephens, 2007) or on charters as a pictorial invocation (*e.g.*, British Library MS Cotton Augustinus ii 4 (Sawyer, 1968, no. 114); British Library MS Cotton Augustinus ii 93 (Sawyer, 1968, n. 163; British Library MS Cotton Vespasian A viiii, *f.* 4r (Sawyer 1968, no. 745)). Coins also featured various cross designs: not only a vertical cross, but also a diagonal cross, and a cross with wedges in angles (Figure 2). Crosses were seen not only in the vertical plane, but also in the horizontal plane, in the cases of place-names ("crossing") and the plans of churches (Rumble, 2006, p. 29; Hinton, 2010, p. 334, Figure 9.2). As is apparent from these examples, we should be aware of this visual culture of the Anglo-Saxons when considering the actual shape of the cross.

**Appendix 4: The Interpretation of *æfter***

**Appendix 4.1: The Interpretation of *æfter* in Old English Dictionaries**

In the discussion on the observational timing, N15a and N15b state "in Old English (and Latin) the word after or *æfter* (in Latin: *post*) can also mean during (sunset)" (N15a, p. 239) and "Both the English *after/æfter* (Clark Hall 1960) and the Latin *post* (Niermeyer 1976) had the meaning of both our todays after as well as during/around [*sic*]" (N15b, p.918), respectively. Consulting the dictionary by Niermeyer (1976), we find the entry of "post" on pp.817-818. The English meanings given in Niermeyer (1976, pp. 817-818) are: "1. towards", "2. after, in search of", "3. with, by, at", "4. in the hand of, in the power of", and "5. deputizing for". Despite our thorough search, we could not find the meaning of "during" or "around" in Niermeyer (1976) that N15b relied on. We also consulted the *Dictionary of Medieval Latin from British Sources* (Howlett *et al*., 2009, pp.2361-2362), which specializes in the usage of Medieval Latin in the British Isles and confirmed the entry of *post* does not give "during" or "around" but "after", as in Classic Latin. Therefore, we conclude that the Latin variants of the Anglo-Saxon Chronicle seem to reject the interpretation of "during/around sunset" as suggested by N15a and N15b, as long as we read them literally.

Limited philological insight might delude us to confirm their interpretation for *æfter* of *æfter sunnan setlgonge/setlgange* in Old English. As shown above, N15b (p. 918) cite Clark Hall (1960) for this interpretation as "during sunset". This dictionary was first published in 1894 and has been reprinted for a long time with its title of *A Concise Anglo-Saxon Dictionary*. The entry of *æfter* of Clark Hall (1960, p. 5) shows its first meaning as "(local and temporal) after, along, behind, through, throughout, during". At least, we can find "during" in this entry.

Nevertheless, the entry states this meaning is both in the "local and temporal" context. As "after sunset" was used in the context of time, we need to extract the meaning in the "temporal context". Moreover, Clark Hall's dictionary is designed for learners, contrary to the scholarly dictionaries such as Bosworth-Toller and the DOE (*Dictionary of Old English*) as reviewed in Fulk (2009) and Rodríguez (2018).

Therefore, we consider it would be more appropriate to consult these scholarly dictionaries for scientific discussions, rather than dictionaries for learners. The Dictionary by Bosworth and Toller (1921, p. 10) interprets this preposition as, "1. local and temporal *dat.*— after; post:" and "2.

extension over space or time,—Along, through, during; κατα, per", in the context of time. As long as relying upon the dictionary by Bosworth and Toller (1921, p.10), *æfter* with the meaning of "during" corresponds not with *post* but with *per* in Latin. Consideration of both the MS F (f.48v.) and the Chronicle of Æthelweard (Campbell, 1962, pp. 25-26) shows their Latin variants not as *per solis occubitum/occasum* but *post solis occubitum/occasum*; hence the phrase *æfter sunnan setlgonge/setlgange* cannot be interpreted as "during sunset".

DOE no longer provides the meaning of "during" in its entry for *æfter*[4]. When we see the entry of *æfter* as the preposition of "II.C. referring to time", its meanings are given respectively as: "II.C.1. following (someone / something) in succession, succeeding, after"; "II.C.2. after an interval of, following the passage of (a period of time)"; "II.C.3. subsequent to a point in time (day, hour, season, etc.)"; "II.C.4. subsequent to an event"; "II.C.6. in combination with the demonstrative pronoun and (usually) the particle þe, forming a conjunction (only in prose, freq. in Or; cf. senses II.B.3 and II.E.7)"; "II.C.5. with dative, instrumental or accusative of the demonstrative pronouns (se, þes), forming an adverb phrase"; and "II.C.7. subsequent to and in consequence of, as a result of, because of". The meaning of "during" is not even given. Other meanings through, throughout, or along are related not with "II.C. referring to time" but "II.A. referring to place" (II.A.2-4). Therefore, again, the latest philological results on Old English favour not the interpretation during sunset but after sunset. In short, the philological results tell us to interpret *æfter sunnan setlgonge/setlgange* in Old English or *post solis occubitum/occasum* in Medieval Latin as after sunset literally and not to manipulate them to mean during sunset.

**Appendix 4.2: The Interpretation of *æfter* in the Alfred's Law Code**

The Alfred's Law Code (El.25; Treschow, 1994, p.93) reads *Gif ðeof brece mannes hus nihtes 7 he weorðe Bær ofslegen, ne sie he na mansleges scyldig. Gif he siððan æfter sunnan upgonge. Bis deð, he bið mansleges scyldig 7 he ðonne self swelte, buton he nieddæda wære* which would be translated as "If a thief breaks into a man's house by *night* and he is slain there, he is not guilty of homicide. If he does so *after sunrise*, he is guilty of homicide and he himself shall die, unless he acted out of necessity" (Treschow, 1994, p. 93). If we confine *æfter sunnan upgonge* (after sunrise) as some short time after sunrise, the code cannot cover the whole day together with *nihtes* (by night). Therefore, it would be more realistic to understand "after sunrise" in this code as covering the whole daytime after sunrise.

---

[4] *The Dictionary of Old English: A to I Online*, s.v. *æfter* (https://tapor.library.utoronto.ca/doe/; accessed: 17 November 2018)

Hayakawa et al., 2019, *Solar Physics*, preprint.
The Celestial Sign in the Anglo-Saxon Chronicle in the 770s

**Appendix 5: Records for the Halo Displays in 806**

**Appendix 5.1: Records and their Translations**

MS F, f.51r. [806] .dcccvi. ...

1.1. *Old English Version*: Eac on ðys ylcan geare, .ii. nonas Iunii, rodetacn wearð ateowe[d] on ðan monan anes Wodnesdæges innan ðare dagenge. An eft on ðis geare an kalendas Septembris an wunderlic tre[nd]el wearð ateowed abutan ðare sunnan. ...

1.2. *English Translation*: [Swanton, 2000, p.59]

MS F 806. [As E]....Also in this same year, on 4 June, the sign of the cross appeared in the moon one Wednesday at the dawning: and again in this year, on 30 August, an amazing ring appeared around the sun.

2.1. *Latin Version*: Hoc anno etiam, .ii. nonas Iunii, luna quarta decima, signum crucis mirabili modo in luna apparuit feria .v. aurora incipiente, hoc modo #. Eodem anno, .iii. kalendas Septembris, luna xii., die dominica, hora. .iiii., corona mirabilis incircuitu solis apparuit. (v. 8, p. 59)

2.2. English Translation: Also in this year, on 4th June, the 14th day of the Moon, the sign of the cross in a remarkable fashion, appeared in the Moon on Wednesday in dawn, like this #. In the same year, on 30 August, the 12th day of the Moon, on Sunday, at the fourth hour, a wonderful crown appeared around the Sun.

**Appendix 5.2: The Drawing in the ASC MS F and the Continental Chronicles**

The halo records in 806 are found only in MS F, whereas parallel records are not found in other manuscripts. The reason for this isolated attestation is due to the fact that the scribe of MS F had access to a manuscript based on a now lost Latin chronicle from Winchester, which in turn was possibly derived from a continental source. In particular, the text and the drawing of a "cross" appear in exactly the same way as in two continental annals called *Annales Sancti Maximi Trevirensis* (Pertz, 1841, p. 6; *c*. 840) and *Annales Laudensis et Sancti Vincentii* (Pertz, 1888, p. 1294; *c*. 875) (Baker, 2000, pp. xlvi-xlvii, 59). Therefore, the observers and the observational sites were probably different between the records in 776 (observed in Britain) and 806 (observed in Continental Europe), and the drawing in the ASC MS F itself is also not an original one based on the original observation but a copy from continental chronicles via the lost Winchester Chronicle. Here, MS F (f.51r.) explicitly reported the existence of the Sun and the Moon as "a cross appeared *in the moon*" and "a wonderful crown was seen *around the sun*". The observer who contributed to the MS F seemed aware of the relationship between the Sun/Moon and a halo display. On the other hand, as for the "red cross" in 776, the observers described this event as *read Cristes mel* (red sign of Christ) and related this phenomenon neither with the Sun nor with the Moon in any of the variants.

**Appendix 6: Modern Instrumental Observations for Auroral Displays during Twilight**

The previous eyewitness records are consistent with modern all-sky auroral observations. Figure 8 summarizes all-sky auroral images taken at South Pole Station (S 90°) (Ebihara *et al*., 2007). The magnetic south (magnetically poleward) is to the left-bottom direction. Sunset occurs for the first time in each year at South Pole Station around 21-22 March because this station is on the Earth's rotation axis. At 02:16 UT on 7 April, the solar elevation angle is −6.6°, which refers to nautical twilight. The scattered sunlight occupies the large area, but the auroral structures can be distinguished. At 10:42 UT on 11 April, the solar elevation angle decreases to −8.2°, which also belongs to the nautical twilight. The scattered sunlight is weakened. The auroral structures are clearly visible.

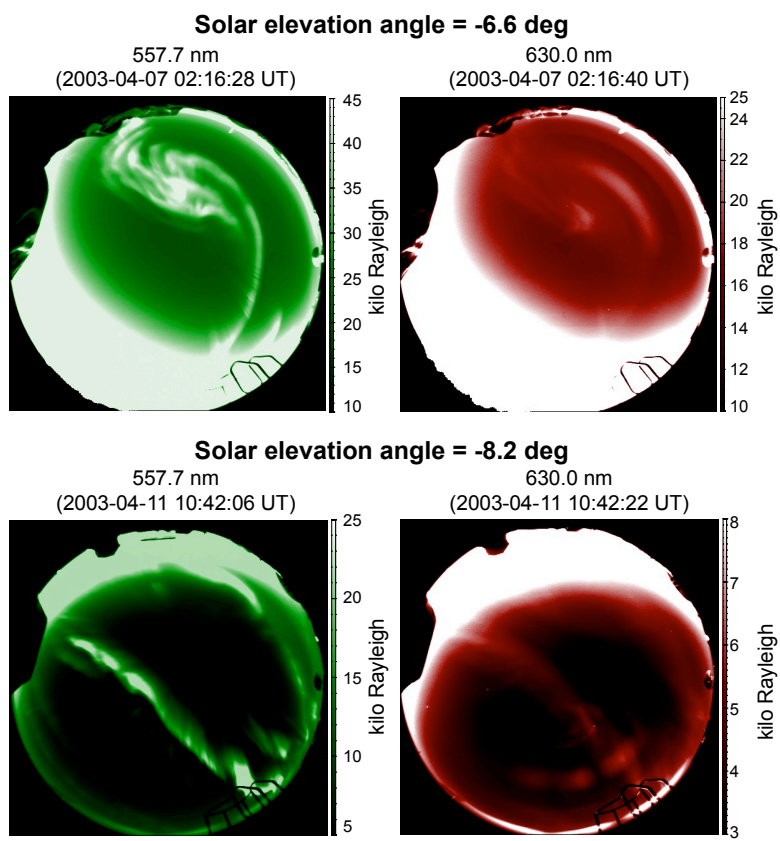

Figure 8: All-sky auroral images at South Pole Station taken at (top) 02:16 UT on 7 April 2003 and (bottom) 10:42 UT on 11 April 2003. The left panels show the images at 557.7 nm (green) and the right ones show the images at 630.0 nm (red). The auroral image data from the South Pole Station is available at the website: http://www.southpole-aurora.org.

Therefore, these images imply that an auroral display is visible even during twilight if its brightness is sufficient. Ebihara *et al*. (2017) consider the cause of auroral displays with extreme brightness (IBC, class IV) in low to mid magnetic latitudes as being due to the precipitation of high-intensity low-energy electrons. This evidence shows that disproving the possibility of an aurora because of "the timing after sunset usually means twilight" (criterion 1 in N15a) contradicts the observational evidence.